\documentclass[journal]{IEEEtran}

\ifCLASSINFOpdf

\else
 
\fi

\usepackage{cite}
\usepackage{amsmath,amssymb,amsfonts}
\usepackage{textcomp}
\usepackage{subfigure}
\usepackage{stfloats}
\usepackage{multirow}
\usepackage{algorithm}
\usepackage{algorithmic}
\usepackage{graphicx}
\usepackage{setspace}
\usepackage{color}
\usepackage{bm}
\usepackage{mathrsfs}
\usepackage{amssymb}
\usepackage{epstopdf}
\usepackage{ntheorem} 
 
\newtheorem{assumption}{Assumption} 
\newtheorem{theorem}{Theorem} 
 
\newtheorem{Proposition}{Proposition}

\newenvironment{proof}{{\indent \indent \it Proof:}}{\hfill $\square$\par}

\begin{document}
\title{
Wavefront Transformation-based Near-field\\
Channel Prediction for Extremely\\
Large Antenna Array with Mobility}
\author{Weidong~Li,~Haifan~Yin,~\IEEEmembership{Senior Member,~IEEE},~Ziao~Qin,~and~M\'{e}rouane~Debbah,~\IEEEmembership{Fellow,~IEEE}
\thanks{W. Li, H. Yin and Z. Qin are with Huazhong University of Science and Technology, 430074 Wuhan, China (e-mail: weidongli@hust.edu.cn, yin@hust.edu.cn, ziao\_qin@hust.edu.cn).}
\thanks{M. Debbah is with KU 6G Research Center, Khalifa University of Science and Technology, P O Box 127788, Abu Dhabi, UAE (email: merouane.debbah@ku.ac.ae) and also with CentraleSupelec, University Paris-Saclay, 91192 Gif-sur-Yvette, France.}
\thanks{The corresponding author is Haifan Yin.}
\thanks{This work is supported by the National Natural Science Foundation of China under Grant 62071191.}}

\markboth{}%
{Shell \MakeLowercase{\textit{et al.}}: Bare Demo of IEEEtran.cls for IEEE Journals}

\maketitle

\begin{abstract}
This paper addresses the mobility problem in extremely large antenna array (ELAA) communication systems. 
In order to account for the performance loss caused by the spherical wavefront of ELAA in the mobility scenario, we propose a wavefront transformation-based matrix pencil (WTMP) channel prediction method. 
In particular, we design a matrix to transform the spherical wavefront into a new wavefront, which is closer to the plane wave. 
We also design a time-frequency projection matrix to capture the time-varying path delay. 
Furthermore, we adopt the matrix pencil (MP) method to estimate channel parameters.
Our proposed WTMP method can mitigate the effect of near-field radiation when predicting future channels.
Theoretical analysis shows that the designed matrix is asymptotically determined by the angles and distance between the base station (BS) antenna array and the scatterers or the user when the number of BS antennas is large enough.  
For an ELAA communication system in the mobility scenario, we prove that the prediction error converges to zero with the increasing number of BS antennas.
Simulation results demonstrate that our designed transform matrix efficiently mitigates the near-field effect, and that our proposed WTMP method can overcome the ELAA mobility challenge and approach the performance in stationary setting. 
\end{abstract}


\IEEEpeerreviewmaketitle

\section{Introduction}
\IEEEPARstart{T}{he} fifth generation (5G) mobile communication systems attain superior spectral and energy efficiencies by introducing the massive multiple-input multiple-output (MIMO) technology \cite{Marzetta10TCom}. 
Compared to 5G, the future sixth generation (6G) wireless communication systems is expected to achieve high throughput by utilizing some new promising technologies, e.g., extremely large antenna array (ELAA) \cite{Cheng19TWC}, Terahertz communications \cite{Wang23TVT}, and reconfigurable intelligent surface (RIS) \cite{Huang19TWC}. 

The ELAA deploys enormous antennas, significantly increasing the array aperture \cite{Podkurkov21TSP, Lu23TCom, Molisch07, Lu22TWC}.
The radiative fields of the array contain the near field and far field, and the boundary between the two fields is Rayleigh distance, defined as $\frac{{2{D^2}}}{\lambda }$ with $D$ denoting array aperture and $\lambda $ being wavelength \cite{Molisch07}.
The near-field region expands with the increasing array aperture, bringing in near-field effects that significantly impact the channel conditions.
The user equipment (UE) or the scatterers are even located within the near-field region, which makes the conventional plane wavefront assumption invalid \cite{Lu22TWC}.

Considering the near-field effects of ELAA, the spherical wavefront assumption can model the near-field channel more accurately \cite{Yang21TWC}.
The advantage of ELAA on the spectral efficiency (SE) relies on the accurate channel state information (CSI). 
Recently, several works have studied ELAA channel estimation.
\cite{21TSPBayesian} proposes a Bayesian channel estimation scheme, and points out that the discrete Fourier transform (DFT) matrix may not perform as expected in the near-field channel because of the unattainable angular-domain sparsity.  
The work in \cite{20WCLChannel} supposes the ELAA has directionality, and estimates the ELAA channel by orthogonal matching pursuit (OMP) algorithm. 
The above works aim to estimate the exact ELAA channel. However, the spherical wavefront model (SWM) is very complex and needs to be simplified \cite{Magoarou19}. 

To simplify the SWM, the authors in \cite{96Validity} approximate the spherical wavefront as a parabolic wavefront, which is more accurate than the plane wavefront and less complex than the spherical wavefront.
The approximation radiative region is called the “Fresnel region” with a range of $\left[ {0.62\sqrt {\frac{{{D^3}}}{\lambda }} ,\frac{{2{D^2}}}{\lambda }} \right]$.
\cite{19TAP} first calculates the fourth-order cumulant of the array to decouple the distance and angles, and then separately estimates the parameters based on the multiple signal classification (MUSIC) algorithm. 
Other subspace-based methods, e.g., estimation of signal parameters via rational invariance techniques (ESPRIT), can also estimate channel parameters when the angles and distance are decoupled by the channel covariance matrix \cite{22RIS}.
However, the above estimation algorithms are high-complexity.
Some neural network (NN) algorithms, e.g., complex-valued neural network (CVNN) \cite{20TVT} and convolutional neural network (CNN) \cite{23WCL}, are trained to estimate the near-field channel. 
Yet, the generalization of NN algorithms still needs to be enhanced. 
Different from the conventional DFT matrix, \cite{22TComDai} designs a polar-domain transform matrix containing angle and distance information to describe channel sparsity. 
By exploiting the polar-domain sparsity, the authors design an OMP-based method to achieve more accurate channel estimation. 
However, the above literature does not consider the mobility problem.

The mobility problem (or “curse of mobility”) \cite{Yin20JSAC,21GaoJSAC} is one typical problem that degrades the performance of massive MIMO.
The UE movement and CSI delay are two main reasons causing the performance decline.
Specifically, the UE movement makes the channel fast-varying, and a large CSI delay causes the estimated CSI to be outdated, making the precoder unusable. 
Channel prediction is an effective solution to the mobility problem. 
The authors in \cite{Yin20JSAC} propose a Prony-based angular-delay domain (PAD) channel prediction method, which is asymptotically error-free when the number of antennas is large enough. 
With the 2D-DFT matrices, the PAD method exploits the angular-delay-domain sparsity. 
However, the ELAA communication system introduces an extra parameter, i.e., distance, and the DFT matrix cannot describe the angular-domain sparsity.
Additionally, the movement of UEs introduces the time-varying path delays. 
The discrete prolate spheroidal (DPS) sequence can capture the slightly varying path delay in a WiFi communication scenario \cite{12TomasTVT}. 
However, in the mobility environment, the path delay may vary substantially, which causes the DPS sequence not to achieve the expected performance.
Therefore, the existing channel prediction methods are unsuitable under the spherical wavefront assumption in the ELAA channel.

In order to fill the above gaps and address the mobility problem of the ELAA channel in this paper, we propose a novel wavefront transformation-based matrix pencil (WTMP) channel prediction method. 
Notice that the steering vectors of the near-field channel and far-field channel share the same angles, and the steering vector of the near-field channel contains an extra distance parameter. 
The key idea is designing a matrix to transform the spherical wavefront and make it closer to the plane wave.
In such a way, the near-field effects may be mitigated.
In the literature, several works have designed methods to transform the near-field estimation to the far-field estimation, e.g., exploiting the fourth-order cumulant of the array \cite{19TAP} and calculating the channel covariance matrix \cite{22RIS}.
Different from the existing methods that aims to simplify the near-field parameters estimation, our proposed WTMP method transforms the near-field channel to the far-field channel.

In this paper, by utilizing the OMP algorithm, we first estimate the channel parameters, i.e., the number of paths, distance, elevation angle of departure (EOD), and azimuth angle of departure (AOD). 
Then, based on the estimated parameters, we design a wavefront-transformation matrix containing the angles and distance information.
Next, to capture the time-varying path delay, we design a time-frequency projection matrix containing the time-varying path delay information. 
The designed matrix is a block matrix, with each sub-block matrix containing the Doppler and path delay information at a certain moment.
The different sub-block matrices are designed based on the Doppler and delay information at different moments.
After that, we project the channel onto the angular-time-frequency domain by constructing an angular-time-frequency projection matrix that consists of the designed wavefront-transformation matrix, time-frequency projection matrix, and DFT matrix. 
Finally, we adopt the matrix pencil (MP) method to estimate the Doppler using the angular-time-frequency-domain CSI.
To the best of our knowledge, our proposed WTMP method is the first attempt to transform the spherical wavefront and predict the ELAA channel.

The contributions of this paper are summarized as follows:

\begin{itemize}
\item 
We propose a WTMP prediction method to address the mobility problem with time-varying path delay in the ELAA channel by designing a wavefront-transformation matrix.
Without straightly estimating the near-field channel, our designed matrix transforms the complex near-field channel estimation into the far-field channel estimation.
The simulations show that our WTMP method significantly outperforms the existing method.

\item We prove that the designed transform matrix depends on the elevation, angle, azimuth angle, and distance between the BS antenna and the scatterers or the UE, as the number of the base station (BS) antennas is large enough. Therefore, the transform matrix can be constructed with estimated angles and distance. 

\item We analyze the asymptotic performance under enough channel samples and a finite number of the BS antennas, and prove that the WTMP method is asymptotically error-free for an arbitrary CSI delay.

\item We further prove that if the number of the BS antennas is large enough and only finite samples are available, the prediction error of our WTMP method asymptotically converges to zero for an arbitrary CSI delay.

\end{itemize}

This paper is organized as follows: We introduce the channel model in Sec. \ref{sec:system model}. Sec. \ref{sec:WTMP method} describes our proposed WTMP channel prediction method. The performance of the WTMP method is analyzed in Sec. \ref{sec:performance analysis}. The simulation results are illustrated and discussed in Sec. \ref{sec:simulation}. Finally, Sec. \ref{sec:conclusion} concludes the paper.

Notations: We use boldface to represent vectors and matrices. Specifically, ${{\bf{I}}_m}$, ${{\bf{0}}_{{m_2} \times {m_3}}}$ and ${{\bf{1}}_{{m_4} \times {m_5}}}$ denote $m \times m$ identity matrix, ${{m_2} \times {m_3}}$ zero matrix and ${{m_4} \times {m_5}}$ one matrix. ${({\bf{X}})^T}$, ${({\bf{X}})^*}$, ${({\bf{X}})^H}$, ${({\bf{X}})^\dag }$ and ${({\bf{X}})^{ - 1}}$ denote the transpose, conjugate, conjugate transpose, Moore-Penrose pseudo inverse and inverse of a matrix ${\bf{X}}$, respectively. 
$\delta ( \cdot )$ is Dirac's delta function.
$\mathcal{F}$ denotes the Fourier transform operation.
${\left\| \cdot \right\|_2}$ stands for the $L_2$ norm of a vector, and ${\left\|  \cdot  \right\|_F}$ means the Frobenius norm of a matrix. 
$r \{  \cdot \}$ represents the rank of a matrix.
${\rm{diag}}\{  \cdot \} $ denotes the diagonal operation of a matrix. $E\{  \cdot \} $ is the expectation operation, and $eig( \cdot )$ denotes the eigenvalue decomposition operation (EVD). $angle( \cdot )$ takes the angle of a complex number.
$ \langle {\bf{x}}, {\bf{y}} \rangle $ represents the inner product of vector ${\bf{x}}$ and ${\bf{y}}$. ${\bf{X}} \otimes {\bf{Y}}$ is the kronecker product of ${\bf{X}}$ and ${\bf{Y}}$. $\mathop  = \limits^\Delta  $ is used to define a new formula.

\section{Channel Model}\label{sec:system model}

\par We consider a TDD massive MIMO system where the BS deploys an ELAA to serve multiple UEs. 
The BS estimates CSI from the UL pilot. The DL CSI is acquired from the BS by utilizing channel reciprocity \cite{Cheng19TWC}.

\begin{figure}[!htb]
\centering
\includegraphics[width=3.4in]{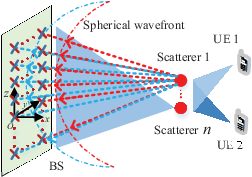}
\caption{The typical UL near-field channel of ELAA communication system.}
\label{fig:1}
\end{figure}

Fig. 1 depicts the near-field channel between the BS and the UE. The BS has a uniform planar array (UPA) consisting of $N_h$ columns and $N_v$ rows. 
The UE has two antenna elements with $-45^\circ $ and $45^\circ $ polarization angles.
The BS is equipped with ${N_t} = N_h N_v$ antennas.
Assume $N_h$ and $N_v$ are even.
The horizontal and vertical apertures of the BS array are $D_h$ and $D_v$.
In the TDD mode, the UL and DL channels share the same bandwidth $B$, which consists of $N_f$ subcarriers with spacing $\Delta f$. 
The channel has $P$ propagation paths, and each path has certain EOD, AOD, delay, distance, Doppler and amplitude.

For the $p$-th path, we denote the elevation angle of arrival (EOA), azimuth angle of arrival (AOA), EOD and AOD as ${\theta _{p,{\rm{EOA}}}}$, ${\phi _{p,{\rm{AOA}}}}$, ${\theta _{p,{\rm{EOD}}}}$ and ${\phi _{p,{\rm{AOD}}}}$, respectively. 
The ranges of angles are ${\theta _{p,{\rm{EOA}}}} \in [0,\pi]$, ${\phi _{p,{\rm{AOA}}}} \in (-\pi,\pi]$, ${\theta _{p,{\rm{EOD}}}} \in [0,\pi]$ and ${\phi _{p,{\rm{AOD}}}} \in (-\pi,\pi]$.
Denote the spherical unit vector of the UE antenna by ${\bf{\hat r}}_p^{\rm{rx}}$:
\begin{equation}
{\bf{\hat r}}_p^{\rm{rx}} = \left[ \begin{array}{l}
\sin {\theta _{p,{\rm{EOA}}}}\cos {\phi _{p,{\rm{AOA}}}}\\
\sin {\theta _{p,{\rm{EOA}}}}\sin {\phi _{p,{\rm{AOA}}}}\\
\ \ \ \ \ \ \cos {\theta _{p,{\rm{EOA}}}}
\end{array} \right].
\!\label{SphericalVectorUE}
\end{equation}
Let ${\beta _p}$ denote the complex amplitude of the $p$-th path.
The Doppler of the $p$-th path is defined as ${\omega _p} = \frac{{{{({\bf{\hat r}}_p^{\rm{rx}})}^T}{\bf{v}}}}{\lambda }$, where ${\bf{v}}$ is the velocity vector of the UE. 
The wavelength $\lambda $ is defined as $\lambda  = \frac{{\rm{c}}}{f_c}$, where ${\rm{c}}$ is the speed of light and $f_c$ is the central carrier frequency.
Denote the location vector of the UE antenna as ${\bf{\bar d}}_{\nu}^{{\rm{rx}}}$.
The BS antenna array is located on the $ yOz$ plane.
Let the antenna element in the center of the BS antenna array be the $ yOz$ coordinate origin, which is located at $\left[ {0,0,0} \right]^T$.
The location of the $s_h$-th column and the $s_v$-th row of the BS antenna array is
\begin{equation}
{\bf{\bar d}}_{{s_h},{s_v}}^{{\rm{tx}}} = {\left[ {\begin{array}{*{20}{c}}{0,}&{{{d_h}({s_h} - \frac{{{N_h}}}{2})},}&{{{d_v}({s_v} - \frac{{{N_v}}}{2})}}\end{array}} \right]^T},
\!\label{BSLocation}
\end{equation}
where $d_h$ and $d_v$ are horizontal and vertical antenna spacings, respectively. 
The ranges of ${d_h}({s_h} - \frac{{{N_h}}}{2})$ and ${d_v}({s_v} - \frac{{{N_v}}}{2})$ are ${d_h}({s_h} - \frac{{{N_h}}}{2}) \in [-\frac{{{D_h}}}{2}, \frac{{{D_h}}}{2} ]$ and ${d_v}({s_v} - \frac{{{N_v}}}{2}) \in [-\frac{{{D_v}}}{2}, \frac{{{D_v}}}{2} ]$.
For notational simplicity, ${\theta _{p,{\rm{EOD}}}}$ and ${\phi _{p,{\rm{AOD}}}}$ are abbreviated as ${\theta _p}$ and ${\phi _p}$.
The location of the $p$-th scatterer is:
\begin{equation}
{{\bf{\bar d}}} = {r_p}\left[ {\begin{array}{*{20}{c}}{\sin {\theta _{p}}\cos {\phi _{p}},}&{\sin {\theta _{p}}\sin {\phi _{p}},}&{{\cos {\theta _{p}}}}\end{array}} \right]^T,
\!\label{UEScatterLocation}
\end{equation}
where ${r_p}$ is the distance from the central BS antenna element to the $p$-th scatterer with a range of ${r_p} \gg \max ({D_h},{D_v})$. 
Eq. (\ref{UEScatterLocation}) can also denote the location of the $\nu$-th UE, if ${r_p}$ is replaced with ${r_{\nu}}$, where ${r_{\nu}}$ denotes the distance between the central BS antenna element to the ${\nu}$-th antenna of the UE.
Let ${h_{{\nu},{s_h},{s_v}}}(t,{\tau})$ denote the channel impulse response between the ${s_h}$-th column and the ${s_v}$-th row of the BS antenna array and the ${\nu}$-th antenna of the UE, which is modelled as \cite{3GPPR16}
\begin{equation}
\begin{array}{l}
\!\!\!{h_{{\nu},{s_h},{s_v}}}(t,{\tau}) \!= \!\sum\limits_{p = 1}^P {{\beta _p}{e^{j\frac{{2\pi {{({\bf{\hat r}}_p^{\rm{rx}})}^T}{\bf{\bar d}}_{\nu}^{{\rm{rx}}}}}{\lambda }}}{e^{ - j\frac{{2\pi ({r_{p,{s_h},{s_v}}} - {r_p})}}{\lambda }}}} \\
\ \ \ \ \ \ \ \ \ \ \ \ \ \ \ \ \ \ \ \ \ \ \ \ {e^{j2\pi {\omega _p}t}}{\delta (\tau  - {\tau _p}(t))},
\end{array}
\!\label{3GPPModelCIR}
\end{equation}
where ${\tau _p}(t)$ is the delay of the $p$-th path \cite{3GPPR16}
\begin{equation}
{\tau _p}(t) = {\tau _{p,0}} + {k_{{\tau _p}}}t = {\tau _{p,0}} - \frac{{{{({\bf{\hat r}}_p^{{\rm{rx}}})}^T}{\bf{v}}}}{{\rm{c}}}t = {\tau _{p,0}} - \frac{{{\omega _p}}}{{{f_c}}}t,
\!\label{TimeVaryingDelay}
\end{equation}
where ${\tau _{p,0}}$ and ${k_{{\tau _p}}}$ are the initial value and the changing rate of delay.
The time-varying path delay can be viewed as the Doppler effect in the frequency domain.
Notice that different paths have different delays, i.e., ${\tau _p}(t) \ne {\tau _q}(t)$.
To describe the effect of path delays, we may also transform ${\delta (\tau  - {\tau _p}(t))}$ to a phase ${e^{ - j2\pi f{\tau _p}(t)}}$ by using the Fourier transform, where $f$ is the frequency.
Therefore, ${h_{{\nu},{s_h},{s_v}}}(t,{\tau})$ is transformed to the channel frequency response ${h_{{\nu},{s_h},{s_v}}}(t,f)$:
\begin{equation}
\begin{array}{l}
\!{h_{{\nu},{s_h},{s_v}}}(t,f)\! =\! \mathcal{F}({h_{{\nu},{s_h},{s_v}}}(t,{\tau})) \!= \!\sum\limits_{p = 1}^P {\beta _p}{e^{j\frac{{2\pi {{({\bf{\hat r}}_p^{\rm{rx}})}^T}{\bf{\bar d}}_{\nu}^{{\rm{rx}}}}}{\lambda }}}\\
\ \ \ \ \ \ \ \ \ \ \ \ \ \ \ \ \ {e^{ - j\frac{{2\pi ({r_{p,{s_h},{s_v}}} - {r_p})}}{\lambda }}} {e^{ - j2\pi f{\tau _p}(t)}}{e^{j2\pi {\omega _p}t}},
\end{array}
\!\!\!\label{3GPPModel}
\end{equation}
where ${r_{p,{s_h},{s_v}}}$ denotes the distance from the ${s_h}$-th column and the ${s_v}$-th row of the BS antenna array to the $p$-th scatterer:
\begin{equation}
\begin{array}{l}
\!{{r}_{p,{s_h},{s_v}}} \!\!=\!\! {\left\| {{{{\bf{\bar d}}}} - {\bf{\bar d}}_{{s_h},{s_v}}^{{\rm{tx}}}} \right\|_2}\!=\! {r_p}\sqrt {1 \!- \!\frac{2{\Psi ({s_h}, {s_v})}}{{{r_p}}} \!\!+\!\! \frac{{\Upsilon ({s_h}, {s_v})}}{{{r_p^2}}}} 
\end{array},
\!\label{DistanceBSScatter}
\end{equation}
with
\begin{equation}
\Psi ({s_h}, {s_v}) \!=\! \sin {\theta _p}\sin {\phi _p}d_h({s_h} - \frac{{{N_h}}}{2}) \!+\! \cos {\theta _p}d_v({s_v} \!- \frac{{{N_v}}}{2}),
\!\label{DistancePart1}
\end{equation}
and
\begin{equation}
\Upsilon ({s_h}, {s_v})  = {d_h^2}{({s_h} - \frac{{{N_h}}}{2})^2} + {d_v^2}{({s_v} - \frac{{{N_v}}}{2})^2}.
\!\label{DistancePart2}
\end{equation}
Since ${r_p} \gg \max ({D_h},{D_v})$, we may obtain ${r_p} \gg {\Psi ({s_h}, {s_v})}$ and ${r_p} \gg {\Upsilon ({s_h}, {s_v})}$.
With a Fresnel approximation expansion $\sqrt {1 + x}  \approx 1 + \frac{1}{2}x - \frac{1}{8}{x^2}$, we may approximate the distance ${r_{p,{s_h},{s_v}}}$ as
\begin{equation}
\begin{array}{l}
{{ r}_{p,{s_h},{s_v}}^{\rm{near}}} \approx {{ r}_p} - {\Psi ({s_h}, {s_v})} + \frac{{\Upsilon ({s_h}, {s_v})} - {{\Psi ({s_h}, {s_v})^2}}}{{2{{ r}_p}}}.
\end{array}
\!\!\!\label{DistancePart2Fresnel}
\end{equation}
Next, we will determine an approximation region where the error of the distance ${{ r}_{p,{s_h},{s_v}}^{\rm{near}}}$ in Eq. (\ref{DistancePart2Fresnel}) is negligible.
Applying a binomial expansion $\sqrt {1 + {x}}  \approx 1 + \frac{{{x}}}{{2}}$, the distance ${r_{p,{s_h},{s_v}}}$ under the far-field assumption is approximated by:
\begin{equation}
\begin{array}{l}
\!\!\!\!\!{r_{p,{s_h},{s_v}}^{\rm{far}}} \approx {r_p} - \Psi ({s_h}, {s_v}).
\end{array}
\!\!\!\label{DistancePart2Farfield}
\end{equation}
With a three-order Taylor expansion $\sqrt {1 + x}  \approx 1 + \frac{1}{2}x - \frac{1}{8}{x^2} + \frac{1}{{16}}{x^3} + \mathcal{O}({x^4})$, the distance ${r_{p,{s_h},{s_v}}}$ is approximated by 
\begin{equation}
\begin{array}{l}
{{ r}_{p,{s_h},{s_v}}} \approx {{ r}_p} - {\Psi ({s_h}, {s_v})} + \frac{{\Upsilon ({s_h}, {s_v})} - {{\Psi ({s_h}, {s_v})^2}}}{{2{{ r}_p}}}\\
  \ \ \ \ \ \ \ \ \ \ +\frac{{{{\Psi ({s_h}, {s_v})}}{\Upsilon ({s_h}, {s_v})} - {\Psi ({s_h}, {s_v})}^3}}{{2{{ r}_p^2}}}.
\end{array}
\!\!\!\label{DistancePart2ThreeOrerTaylor}
\end{equation}
Denote the phases of the exact spherical wavefront, the approximated near-field spherical wavefront and the far-field plane wavefront as ${\Phi _{p,{s_h},{s_v}}} = \frac{{2\pi }}{\lambda }{r_{p,{s_h},{s_v}}}$, $\Phi _{p,{s_h},{s_v}}^{{\rm{near}}} = \frac{{2\pi }}{\lambda }r_{p,{s_h},{s_v}}^{{\rm{near}}}$, and $\Phi _{p,{s_h},{s_v}}^{{\rm{far}}} = \frac{{2\pi }}{\lambda }r_{p,{s_h},{s_v}}^{{\rm{far}}}$, respectively.
Therefore, the phase discrepancy between the spherical wavefront and the approximated near-field spherical wavefront is calculated by
\begin{equation}
\!\!\begin{array}{l}
\!\!\!\!{\Delta ^{{\rm{near}}}} \!\!=\!\! \left| {{\Phi _{p,{s_h},{s_v}}} \!\!-\!\! \Phi _{p,{s_h},{s_v}}^{{\rm{near}}}} \right| \approx \frac{{2\pi }}{\lambda }\left|\frac{{{{\Psi ({s_h}, {s_v})}}{\Upsilon ({s_h}, {s_v})} - {\Psi ({s_h}, {s_v})}^3}}{{2{{ r}_p^2}}} \right|.
\end{array}
\!\!\!\!\!\label{PhaseDiscrepancyNearfield}
\end{equation}
Since ${\phi _p} \in ( - \pi ,\pi ]$, $ - \frac{{{D_h}}}{2} \le {d_h}({s_h} - \frac{{{N_h}}}{2}) \le \frac{{{D_h}}}{2}$, and $ - \frac{{{D_v}}}{2} \!\le\! {d_v}({s_v} - \frac{{{N_v}}}{2}) \!\le\! \frac{{{D_v}}}{2}$, we may obtain that if ${\phi _p} = \frac{{{\pi}}}{2}$, ${d_h}({s_h} - \frac{{{N_h}}}{2}) = \frac{{{D_h}}}{2}$ and ${d_v}({s_v} - \frac{{{N_v}}}{2}) = \frac{{{D_v}}}{2}$, the maximum phase discrepancy may be achieved as ${\Delta _{\max } ^{{\rm{near}}}} \approx \frac{\pi {{{\xi}_{\max }}({\theta _p})}}{8{\lambda {{ r}_p}^2}}$, where 
\begin{equation}
\begin{array}{l}
\!\!\!\!\!\!\xi({\theta _p}) = {( {\sin {\theta _p}{D_h} + \cos {\theta _p}{D_v}} ){{( {\cos {\theta _p}{D_h} - \sin {\theta _p}{D_v}} )}^2}},
\end{array}
\!\!\!\!\!\!\!\label{PhaseDiscrepancyNearfieldF}
\end{equation}
and ${{\xi_{\max }}({\theta _p})}$ is the maximum value of ${\xi({\theta _p})}$ for a range of ${\theta _{p}} \in [0,\pi]$.
The boundary between the approximated spherical wavefront and the exact spherical wavefront is determined by the condition that the largest phase discrepancy is no more than $\frac{{\pi }}{8}$ \cite{Lu23TCom}, i.e., ${\Delta _{\max} ^{{\rm{near}}}} \le \frac{{\pi }}{8}$. Therefore, we may obtain:
\begin{equation}
\begin{array}{l}
\sqrt {\frac{{{{{\xi_{\max }}({\theta _p})}}}}{\lambda }}  \le {{ r}_p}.
\end{array}
\!\!\!\label{PhaseDiscrepancyNearfieldDistance}
\end{equation}
Similarly, we calculate the phase discrepancy between the approximated near-field spherical wavefront and the far-field plane wavefront as
\begin{equation}
\begin{array}{l}
\!\!\!\!\Delta ^{{\rm{far}}} = \left| {\Phi _{p,{s_h},{s_v}}^{{\rm{near}}} \!\!- \Phi _{p,{s_h},{s_v}}^{{\rm{far}}}} \right| \approx \frac{{2\pi }}{\lambda } \left| \frac{{\Upsilon ({s_h}, {s_v})} - {{\Psi ({s_h}, {s_v})^2}}}{{2{{ r}_p}}} \right|.
\end{array}
\!\!\!\label{PhaseDiscrepancyFarfield}
\end{equation}
When ${\theta _p} = \frac{{{\pi}}}{2}$, ${\phi _p} = 0$, ${d_h}({s_h} - \frac{{{N_h}}}{2}) = \frac{{{D_h}}}{2}$ and ${d_v}({s_v} - \frac{{{N_v}}}{2}) = \frac{{{D_v}}}{2}$, we obtain the maximum phase discrepancy as $\Delta _{\max } ^{{\rm{far}}} \approx \frac{{\pi }({{D_h}^2 + {D_v}^2})}{4{{\lambda }{{ r}_p}}}$.
Let the phase discrepancy $\Delta _{\max } ^{{\rm{far}}}$ more than $\frac{{\pi }}{8}$: $\Delta _{\max } ^{{\rm{far}}}  >  \frac{{\pi }}{8}$ \cite{Lu23TCom}. We may obtain:
\begin{equation}
\begin{array}{l}
{{ r}_p} < \frac{{2({D_h^2} + {D_v^2})}}{\lambda }.
\end{array}
\!\!\!\label{PhaseDiscrepancyFarfieldDistance}
\end{equation}
Eventually, the approximation region is determined by $[\sqrt {\frac{{{{{\xi_{\max }}({\theta _p})}}}}{\lambda }}, \frac{{2({D_h^2} + {D_v^2})}}{\lambda }]$, where the error of the distance ${{ r}_{p,{s_h},{s_v}}^{\rm{near}}}$ is negligible.

The 3-D near-field steering vector ${\bf{a}}({\theta _p},{\phi _p},{r_p}) \in {{\mathbb{C}}^{N_t \times 1}}$ containing the distance, EOD, and AOD is
\begin{equation}
\begin{array}{l}
{\bf{a}}({\theta _p},{\phi _p},{r_p}) = {\rm{diag}}\left\{ {{{\bf{a}}}({r_p})} \right\}({{\bf{a}}_h}({\theta _p},{\phi _p}) \otimes {{\bf{a}}_v}({\theta _p})),
\end{array}
\!\label{SWMSteeringVector}
\end{equation}
where ${{{\bf{a}}}({r_p})} \in {{\mathbb{C}}^{{N_t} \times 1}}$ is a distance response vector:
\begin{equation}
{{\bf{a}}}({r_p}) = {\left[ {\begin{array}{*{20}{c}}{1,}&{ \cdots ,}&{{e^{ - j\frac{{2\pi }}{\lambda }\frac{{\Upsilon ({N_h}, {N_v})} - {{\Psi ({N_h}, {N_v})^2}}}{{2{{ r}_p}}}  }  }  } \end{array}} \right]^T}.
\!\label{DistanceVectorH}
\end{equation}
The two matrices ${{\bf{a}}_h}({\theta _p},{\phi _p}) \in {{\mathbb{C}}^{{N_h} \times 1}}$ and ${{\bf{a}}_v}({\theta _p}) \in {{\mathbb{C}}^{{N_v} \times 1}}$ are expressed as: 
\begin{equation}
{{\bf{a}}_h}({\theta _p},{\phi _p}) = {\left[ {\begin{array}{*{20}{c}}{1,}&{ \cdots ,}&{{e^{j\frac{{2\pi }}{\lambda }\sin {\theta _p}\sin {\phi _p}{d_h}({N_h} - 1)}}} \end{array}} \right]^T},
\!\label{PWMSteeringVectorH}
\end{equation}
and
\begin{equation}
{{\bf{a}}_v}({\theta _p}) = {\left[ {\begin{array}{*{20}{c}}{1,}&{ \cdots ,}&{{e^{j\frac{{2\pi }}{\lambda }\cos {\theta _p}{d_v}({N_v} - 1))}}} \end{array}} \right]^T}.
\!\label{PWMSteeringVectorV}
\end{equation}
Therefore, the 3-D far-field steering vector is expressed as:
\begin{equation}
{\bf{a}}({\theta _p},{\phi _p}) = {{\bf{a}}_h}({\theta _p},{\phi _p}) \otimes {{\bf{a}}_v}({\theta _p}).
\!\label{PWMSteeringVector}
\end{equation}
Denote the channel between all BS antennas and the ${\nu}$-th UE antenna at time $t$ and frequency $f$ as ${{\bf{h}}_{\nu}}(t,f) \in {{\mathbb{C}}^{1 \times N_t}}$.
The channels at all subcarriers are ${{\bf{H}}_{\nu}}(t) = \left[ {{\bf{h}}_{\nu}^T(t,{f_1}),{\bf{h}}_{\nu}^T(t,{f_2}), \cdots ,{\bf{h}}_{\nu}^T(t,{f_{{N_f}}})} \right]$, where ${f_{{n_f}}} = {f_1} + ({n_f} - 1)\Delta f$ is the ${n_f}$-th subcarrier frequency.
We rewrite ${{\bf{H}}_{\nu}}(t)$ as
\begin{equation}
{{\bf{H}}_{\nu}}(t) = {{\bf{A}}_{\nu}}{{\bf{C}}_{\nu}}{{\bf{B}}_{\nu}}(t),
\!\label{HChannelACB}
\end{equation}
where ${{\bf{C}}_{\nu}} = {\rm{diag}}\left\{ {{c_{{\nu},1}}, \cdots ,{c_{{\nu},P}}} \right\} \in {{\mathbb{C}}^{P \times P}}$ with ${c_{{\nu},p}} = {\beta _p}{e^{j\frac{{2\pi {{({\bf{\hat r}}_p^{rx})}^T}{\bf{\bar d}}_{\nu}^{{\rm{rx}}}}}{\lambda }}}$. 
The matrix ${{\bf{B}}_{\nu}}(t) \in {{\mathbb{C}}^{P \times {N_f}}}$ consists of delay-and-Doppler vectors:
\begin{equation}
{{\bf{B}}_{\nu}}(t) = {\left[ {\begin{array}{*{20}{c}}{{\bf{b}}({\tau _{1,0}},{\omega _1}t),}&{ \cdots ,}&{{\bf{b}}({\tau _{P,0}},{\omega _P}t)} \end{array}} \right]^T},
\!\label{Bu}
\end{equation}
where
\begin{equation}
\begin{array}{l}
\!\!\!\!\!{\bf{b}}({\tau _{p,0}},{\omega _p}t) = \\
\!\!\!\!\!{\left[ \!{{e^{j2\pi ((1 + \frac{{{f_1}}}{{{f_c}}}){\omega _p}t - {f_1}{\tau _{p,0}})}}, \cdots ,{e^{j2\pi ((1 + \frac{{{f_{{N_f}}}}}{{{f_c}}}){\omega _p}t - {f_{{N_f}}}{\tau _{p,0}})}}} \!\right]^T}.
\end{array}\!\!\!\!\!
\!\label{b-tau-Doppler}
\end{equation}
The matrix ${{\bf{A}}_{\nu}} \in {{\mathbb{C}}^{N_t \times P}}$ contains the 3-D near-field steering vectors of all paths:
\begin{equation}
\begin{array}{l}
{{\bf{A}}_{\nu}} = {\left[ {\begin{array}{*{20}{c}}{{\bf{a}}({\theta _1},{\phi _1},{r_1}),}&{ \cdots ,}&{{\bf{a}}({\theta _P},{\phi _P},{r_P})} \end{array}} \right]}\\
 \ \ \ \ \ = ( {\sum\limits_{p = 1}^P {{{\bf{K}}_p}} } ){{\bf{A}}_r}{{\bf{A}}_{\theta ,\phi }},
\end{array}
\!\label{Au-tx}
\end{equation}
where $\sum\limits_{p = 1}^P {{{\bf{K}}_p}}  \in {{\mathbb{C}}^{{N_t} \times {N_t}P}}$ is a block matrix: 
\begin{equation}
\sum\limits_{p = 1}^P {{{\bf{K}}_p}}  = \left[ {\begin{array}{*{20}{c}}{{\bf{I}}_{N_t},}&{{\bf{I}}_{N_t},}&{ \cdots ,}&{{\bf{I}}_{N_t}} \end{array}} \right].
\!\label{KMatrix}
\end{equation}
The diagonal matrix ${{\bf{A}}_r} \in {{\mathbb{C}}^{{N_t}P \times {N_t}P}}$ is composed of the distance response vectors of all paths: 
\begin{equation}
{{\bf{A}}_r} = {\rm{diag}}\left\{ {\begin{array}{*{20}{c}}{{{\bf{a}}^T}({r_1}),}&{ \cdots ,}&{{{\bf{a}}^T}({r_P})} \end{array}} \right\}.
\!\label{A-Distance}
\end{equation}
The matrix ${{\bf{A}}_{\theta ,\phi }} \in {{\mathbb{C}}^{{N_t}P \times P}}$ contains the 3-D far-field steering vectors of all paths:
\begin{equation}
\begin{array}{l}
{{\bf{A}}_{\theta ,\phi }} = \left[ \begin{array}{l}
{\bf{a}}({\theta _1},{\phi _1}),\ \ {{\bf{0}}_{{N_t} \times 1}}, \ \ \cdots ,\ \ {{\bf{0}}_{{N_t} \times 1}}\\
\ \ {{\bf{0}}_{{N_t} \times 1}},\ \ {\bf{a}}({\theta _2},{\phi _2}), \cdots ,\ \ {{\bf{0}}_{{N_t} \times 1}}\\
 \ \ \ \ \ \vdots \ \ \ \ \ \ \ \ \ \ \ \vdots \ \ \ \ \ \ \  \ddots  \ \ \ \ \ \ \ \vdots \\
\ \ {{\bf{0}}_{{N_t} \times 1}},\ \ \ {{\bf{0}}_{{N_t} \times 1}}, \ \ \cdots ,\ {\bf{a}}({\theta _P},{\phi _P})
\end{array} \right]\\
 \ \ \ \ \ \ \ = \left[ {\begin{array}{*{20}{c}}{{\bm{\alpha}} ({\theta _1},{\phi _1}),\!\!\!\!}&{{\bm{\alpha}} ({\theta _2},{\phi _2}),\!\!\!\!}&{\cdots ,\!\!\!\!}&{{\bm{\alpha}} ({\theta _P},{\phi _P})} \end{array}} \right],
\end{array}
\!\label{A-angle}
\end{equation}
where ${\bm{\alpha}} ({\theta _p},{\phi _p})$ is the $p$-th column vector of ${{\bf{A}}_{\theta ,\phi }}$.
The vectorized form of ${{\bf{H}}_{\nu}}(t)$ is given by 
\begin{equation}
\begin{array}{l}
\!\!\!\!\!{{\bf{\bar h}}_{\nu}}{\rm{(}}t{\rm{) \!=\! vec(}}{{\bf{H}}_{\nu}}(t){\rm{)}}{\rm{ = }}\!\!\sum\limits_{p = 1}^P {{c_{u,p}}} {\bf{b}}({\tau _{p,0}},{\omega _p}t) \!\otimes\! {{\bf{a}}}({\theta _p},{\phi _p},{r_p}),
\end{array}
\!\!\!\!\!\label{ChannelVectorized}
\end{equation}
where ${{\bf{\bar h}}_{\nu}}{\rm{(}}t{\rm{)}} \in {{\mathbb{C}}^{{N_t}{N_f} \times 1}}$.

\section{The Proposed WTMP Channel Prediction Method}\label{sec:WTMP method}

In this section, we introduce our proposed WTMP channel prediction method. 
In an ELAA communication system, due to the near-field effects, the spherical wavefront assumption is true in place of the plane wavefront assumption and introduces phase fluctuations among array elements.
To coping with the phase fluctuations challenge, we propose a WTMP method based on the structures of the near-field and far-field channels.
The key to the WTMP method is designing a matrix that transforms the phase of the near-field channel into a new phase. 
Compared to the phase of the near-field channel, the new phase is closer to the one of the far-field channel. 
In general, we first estimate the parameters, i.e., EOD, AOD, and distance, via the OMP algorithm. 
Then, basing on the steering vector estimation of the near-field and far-field channels, we design a wavefront-transformation matrix. 
Next, another time-frequency-domain projection matrix is constructed to track the time-varying path delays. 
Finally, we adopt the MP method to estimate Doppler.
The details will be shown below.

\subsection{The Parameters Estimation}\label{sec:Parameters Estimation}
The near-field channel is still compressible even though the angle sparsity does not hold because the number of paths is usually less than the number of array elements, i.e., $P \le {N_t}$.  
Here we adopt the OMP algorithm to estimate the angles and distance. 
The channels at different subcarriers share the same parameters, i.e., EOD, AOD, and distance. 
For simplification, we use the channel at the first subcarrier to estimate angles and distance. 
The observation channel at the first subcarrier is ${\bf{h}}_{\nu}^T(t,{f_1}) = {{\bf{A}}_{\nu}}{{\bf{C}}_{\nu}}{{\bf{b}}_h}(t,{f_1})$, where 
\begin{equation}
\begin{array}{l}
\!\!\!\!\!{{\bf{b}}_h}(t,{f_1}) = \\
\!\!\!\!\!{\left[ \!\!{\begin{array}{*{20}{c}}{e^{j2\pi (\frac{{({f_c} + {f_1}){\omega _1}t}}{{{f_c}}} - {f_1}{\tau _{1,0}})},\!\!\!\!}&{ \cdots \!,\!\!\!\!}&{{e^{j2\pi (\frac{{({f_c} + {f_1}){\omega _P}t}}{{{f_c}}} - {f_1}{\tau _{P,0}})}}} \end{array}} \!\!\right]}.
\end{array}
\!\!\!\label{Vectorb-t-f1}
\end{equation}
The matrix ${{\bf{A}}_{\nu}}$ may be viewed as a dictionary matrix depending on the tuple $({\theta},{\phi},{r})$. The parameters estimation problem is transformed into a vector reconstructing problem by discretizing the EOA, AOD, and distance with a grid:
\begin{equation}
\begin{array}{l}
\!\!\!\!\!\!\Xi  = \{ ({\theta},{\phi},{r})|{\theta} = {\theta _{\min }},{\theta _{\min }} + \Delta \theta , \cdots ,{\theta _{\max }};\\
\ \ \ \ \ \ \ \ \ \ \ \ \ \ {\phi} = {\phi _{\min }},{\phi _{\min }} + \Delta \phi , \cdots ,{\phi _{\max }};\\
\ \ \ \ \ \ \ \ \ \ \ \ \ \ {r} = {r_{\min }},{r_{\min }} + \Delta r, \cdots ,{r_{\max }}\},
\end{array}
\!\label{Grid}
\end{equation}
where $\Delta \theta$, $\Delta \phi$, and $\Delta r$ are the resolutions of EOD, AOD, and distance. Also, $[{\theta _{\min }},{\theta _{\max }}]$, $[{\phi _{\min }},{\phi _{\max }}]$, and $[{r_{\min }},{r_{\max }}]$ are ranges of EOD, AOD, and distance.
The numbers of sampling grid points ${\bar \theta}$, ${\bar \phi}$, and $\bar r$ are ${M_\theta }$, ${M_\phi }$, and ${M_r}$. 
The dictionary matrix ${{\bf{\bar A}}_{\nu}} \in {{\mathbb{C}}^{{N_h}{N_v} \times {{M_\theta }{M_\phi }{M_r}}}}$ is expressed as
\begin{equation}
{{\bf{\bar A}}_{\nu}} = \left[ {\begin{array}{*{20}{c}}{{\bf{a}}({{\bar \theta }_1},{{\bar \phi }_1},{{\bar r}_1}),}&{ \cdots ,}&{{\bf{a}}({{\bar \theta }_{{M_\theta }}},{{\bar \phi }_{{M_\phi }}},{{\bar r}_{{M_r}}})} \end{array}} \right].
\!\label{DictionaryMatrix}
\end{equation}
Utilizing the OMP algorithm, we may determine a pair of distance and angles in each iteration. 
After $\hat P$ iterations, the number of paths, EOD, AOD, and distance are estimated as $\hat P$, $\bm{\hat \theta} $, $\bm{\hat \phi} $, and $\bm{\hat r}$, respectively.

\subsection{Wavefront Transformation}\label{sec:wavefront transform matrix}
Based on the estimated parameters in Sec. \ref{sec:Parameters Estimation}, we now design a wavefront-transformation matrix in this section. 
For ease of exposition, we first determine the transform matrix based on the mapping relationship between the near-field and far-field steering vectors.
Then, we generate the matrix and determine each entry.
Finally, we design the transform matrix by normalizing each entry. As we focus on the phase fluctuation, only the phase of the generated matrix is needed.

We start by describing the mapping relationship between the near-field and far-field steering vectors.
Denote a matrix ${{\bf{A}}_{{\rm{plane}}}} \in {{\mathbb{C}}^{N_t \times P}}$ containing the 3-D far-field steering vectors of all paths:
\begin{equation}
{{\bf{A}}_{{\rm{plane}}}} = ( {\sum\limits_{p = 1}^P {{{\bf{K}}_p}} } ){{\bf{A}}_{\theta ,\phi }},
\!\label{PWMSteeringVectors}
\end{equation}
where $\sum\limits_{p = 1}^P {{{\bf{K}}_p}}$ and ${{\bf{A}}_{\theta ,\phi }}$ are given in Eq. (\ref{KMatrix}) and Eq. (\ref{A-angle}).
Define a matrix ${{\cal {\bm B}}} \in {{\mathbb{C}}^{N_t \times N_t}}$ describing the mapping relationship between ${{\bf{A}}_{{\rm{plane}}}}$ and ${\bf{A}}_{\nu}$ in Eq. (\ref{Au-tx}):
\begin{equation}
{{{\cal {\bm B}}}{{{\bf{A}}_{\nu}}}} = {{\bf{A}}_{{\rm{plane}}}}.
\!\label{SWMTransformPWM}
\end{equation}
By substituting Eq. (\ref{Au-tx}) and Eq. (\ref{PWMSteeringVectors}) into Eq. (\ref{SWMTransformPWM}),we may rewrite Eq. (\ref{SWMTransformPWM}) as
\begin{equation}
{\bf{G}}{{\bf{A}}_{\theta ,\phi }} = {\bf{0}},
\!\label{SWMTransformPWM0}
\end{equation}
where ${\bf{G}} \in {{\mathbb{C}}^{N_t \times {N_t}P}}$ is expressed as
\begin{equation}
\begin{array}{l}
\!\!{\bf{G}} \!=\! {{\cal {\bm B}}}( {\sum\limits_{p = 1}^P {{{\bf{K}}_p}} } ){{\bf{A}}_r} - \sum\limits_{p = 1}^P {{{\bf{K}}_p}} \!=\! {\left[ {\begin{array}{*{20}{c}}{{{\bf{g}}_1},}&{\cdots ,}&{{{\bf{g}}_{N_t}}} \end{array}} \right ]^T},
\end{array}
\!\!\!\label{Matrix-G}
\end{equation}
and ${{\bf{g}}_n} \in {{\mathbb{C}}^{{N_t}P \times 1}}$, $n = 1, \cdots ,{N_t}$ is expressed as:
\begin{equation}
{{\bf{g}}_n} = {\left[ {\begin{array}{*{20}{c}}{{g_{n,1}},}&{\cdots ,}&{{g_{n,n}},}&{ \cdots ,}&{{g_{n,{N_t}P}}} \end{array}} \right ]^T}.
\!\label{RowVector-G}
\end{equation}
From Eq. (\ref{Matrix-G}), we may notice that matrix ${{\cal {\bm B}}}$ is the wavefront-transformation matrix. 

Then, we will calculate the matrix ${\bf{G}}$.
Perform the SVD of ${{\bf{A}}_{\theta ,\phi }}$: ${{\bf{A}}_{\theta ,\phi }} = {\bf{US}}{{\bf{V}}^H}$, where the unitary matrix ${\bf{U}}$ contains the left singular vectors:
\begin{equation}
{\bf{U}} = \left[ {\begin{array}{*{20}{c}}{{{\bf{u}}_1},}&{ \cdots ,}&{{{\bf{u}}_P},}&{{{\bf{u}}_{P + 1}},}&{ \cdots ,}&{{{\bf{u}}_{{N_t}P}}}\end{array}} \right ],
\!\label{U-A-angle}
\end{equation}
and the rank of ${{\bf{A}}_{\theta ,\phi }}$ is $P$. In other words, ${{\bf{A}}_{\theta ,\phi }} = {{\bf{U}}_{1:P}}{{\bf{S}}_{1:P}}{{\bf{V}}_{1:P}^H}$, where ${{\bf{U}}_{1:P}} = [{{\bf{u}}_1}, \cdots ,{{\bf{u}}_P}]$, the diagonal matrix ${{\bf{S}}_{1:P}}$ contains the first $P$ singular values, and ${{\bf{V}}_{1:P}}$ consists of the first $P$ column vectors of ${\bf{V}}$.
Based on the matrix structure of ${{\bf{A}}_{\theta ,\phi }}$ in Eq. (\ref{A-angle}), the $p$-th column vector of ${\bf{U}}$ is computed by:
\begin{equation}
\begin{array}{l}
{{\bf{u}}_p} = \frac{{{\bm{\alpha}} ({\theta _p},{\phi _p})}}{{{{\left\| {{\bm{\alpha}} ({\theta _p},{\phi _p})} \right\|}_2}}} = {\left[ {\begin{array}{*{20}{c}}{{u_{p,1}},}&{ \cdots ,}&{{u_{p,{N_t}P}}} \end{array}} \right ]^T}, \\
\ \ \ \ \ \ \ \ \ \ \ \ \ \ \ \ \ \ \ \ \ \ \ \ \ \ p = 1, \cdots ,P,
\end{array}
\!\label{UcolumnvectorFirstP}
\end{equation}
and
\begin{equation}
{{\bf{u}}_p} = {\bf{0}}, p = P+1, \cdots ,{N_t}P.
\!\label{UcolumnvectorLast}
\end{equation}
From Eq. (\ref{SWMTransformPWM0}), we may obtain
\begin{equation}
{\bf{G}}{{\bf{U}}_{1:P}}{{\bf{S}}_{1:P}}{{\bf{V}}_{1:P}}^H = {\bf{0}},
\!\label{SWMTransformPWM2}
\end{equation}
where ${\bf{G}}$ is orthogonal to ${{\bf{U}}_{1:P}}$. 
Denote a space $\mathbb{U} = {\rm{span}}\{ {{\bf{u}}_p}:p = 1, \cdots ,P\} $. 
Therefore, the matrix ${\bf{G}}$ falls into the null space of $\mathbb{U}$.
Up to now, the matrix ${\bf{G}}$ is determined.

Next, we will generate the transform matrix ${{\cal {\bm B}}}$ and determine each entry. 
According to Eq. (\ref{Matrix-G}), we may first design matrix ${\bf{G}}$ and then generate matrix ${{\cal {\bm B}}}$.
For matrix ${\bf{G}}$, there are many potential matrices that fall into the null space of $\mathbb{U}$.
Fortunately, we only need to generate a suitable matrix in the null space of $\mathbb{U}$.
For derivation simplicity, we assume that the energy of the $n$-th row vector in matrix ${\bf{G}}$ is concentrated in only one element.
For example, the energy of vector ${{\bf{g}}_n}$ is one, where ${g_{n,n}}$ is close to one, and the other elements are close to zero. 

From Eq. (\ref{SWMTransformPWM2}), it is clear that ${{\bf{g}}_n}$ is orthogonal to all column vectors of ${{\bf{U}}_{1:P}}$. 
Without loss of generality, we assume that the energy of ${{\bf{g}}_n}$ is concentrated in ${g_{n,n}}$ and that the last $({N_t}P - {N_t})$ elements of ${{\bf{g}}_n}$ are zero. 
Since the first ${N_t}$ elements of ${\bf{u}}_p$, $p = 2, \cdots ,P$ are zero, it is easily obtained that ${\bf{u}}_p^H{{\bf{g}}_n} = 0$, $p = 2, \cdots ,P$.
To determine the first ${N_t}$ elements of ${{\bf{g}}_n}$, we may formulate an optimization problem as
\begin{equation}
\max (\left\| {g_{n,n}} \right\|_2),\ {\rm{s.t.}} \ \langle {{\bf{u}}_1},{{\bf{g}}_n} \rangle  = 0,\ \left\| {{{\bf{g}}_n}} \right\|_2^2 = 1.
\!\label{optimization Problem}
\end{equation}
However, it is difficult and very complex to determine ${N_t}$ non-zero entries one by one.
To simplify the optimization problem of Eq. (\ref{optimization Problem}), we assume
\begin{equation}
u_{1,q}^*{g_{n,q}} = {x_1} + i{y_1},\ q = 2, \cdots ,n - 1,n + 1, \cdots ,{N_t},
\!\label{A-C-AllElements}
\end{equation}
\begin{equation}
{g_{n,n}} = {x_2} + i{y_2},
\!\label{A-C-nElements}
\end{equation}
\begin{equation}
{u_{1,n}} = {\eta _1} + i{\kappa _1},
\!\label{u-nElements}
\end{equation}
where ${x_1}$, ${y_1}$, ${x_2}$, ${y_2}$, ${\eta _1}$ and ${\kappa _1}$ are real variables.
Denote the $q$-th element of ${\bm{\alpha}} ({\theta _1},{\phi _1})$ in Eq. (\ref{UcolumnvectorFirstP}) as ${{\alpha}_{1,q}}$. 
Since the last ${N_t}P- {N_t}$ elements of ${\bm{\alpha}} ({\theta _1},{\phi _1})$ are zero, we may obtain ${{\alpha}_{1,q}^*}{{\alpha}_{1,q}} = 1$ and ${{{{\left\| {{\bm{\alpha}} ({\theta _p},{\phi _p})} \right\|}_2}}} = \sqrt{N_t}$.
Eventually, $\left\| {{u_{1,q}}} \right\|_2^2$ in Eq. (\ref{UcolumnvectorFirstP}) is calculated as:
\begin{equation}
\left\| {{u_{1,q}}} \right\|_2^2 = \frac{1}{N_t}, \ q = 1, \cdots ,{N_t}.
\!\label{u-norm}
\end{equation}
Basing on Eq. (\ref{A-C-AllElements}), we may compute
\begin{equation}
{g_{n,q}}g_{n,q}^* = {N_t}(x_1^2 + y_1^2),\ q = 2, \cdots ,n - 1,n + 1, \cdots ,{N_t}.
\!\label{g-norm}
\end{equation}
From $\langle {{\bf{u}}_1},{{\bf{g}}_n} \rangle  = 0$, we readily obtain
\begin{equation}
{g_{n,1}} =  - \frac{{\sum\limits_{q = 2}^{{N_t}} {u_{1,q}^*{g_{n,q}}} }}{{u_{1,1}^*}}.
\!\label{g-n1}
\end{equation}
According to the assumptions and equalities between Eq. (\ref{A-C-AllElements}) and Eq. (\ref{g-n1}), the optimization problem in Eq. (\ref{optimization Problem}) is reformulated as
\begin{equation}
\begin{array}{l}
{\rm{max}}({x_2^2} + {y_2^2}),\\
\ \ \ {\rm{s.t.}} \ \ {({x_2} + \frac{{{z_1}}}{2})^2} + {({y_2} + \frac{{{z_2}}}{2})^2} = \frac{{z_1^2 + z_2^2 + 4{z_3}}}{4},
\end{array}
\!\label{optimization Problem Reformulated}
\end{equation}
where
\begin{equation}
\begin{array}{l}
{z_1} = {N_t}({N_t} - 2)({\eta _1}{x_1} - {\kappa _1}{y_1})\\
{z_2} = {N_t}({N_t} - 2)({\eta _1}{y_1} + {\kappa _1}{x_1})\\
{z_3} = \frac{{1 - {N_t}({N_t} - 1)({N_t} - 2)(x_1^2 + y_1^2)}}{2}.
\end{array}
\!\label{z1-z2-z3}
\end{equation}
Define ${\rm{max}}({x_2^2} + {y_2^2}) = J{({x_1},{y_1})^2}$, where
\begin{equation}
J({x_1},{y_1}) = \sqrt {\frac{{{z_1}^2}}{4} + \frac{{{z_2}^2}}{4}}  + \sqrt {\frac{{{z_1}^2 + {z_2}^2 + 4{z_3}}}{4}} .
\!\label{f-xy}
\end{equation}
Letting $\frac{{\partial J({x_1},{y_1})}}{{\partial (x_1^2 + y_1^2)}} = 0$, we may obtain
\begin{equation}
x_1^2 + y_1^2 = \frac{1}{{{N_t^3} - {N_t^2}}}.
\!\label{x1y1}
\end{equation}
If ${y_2}=0$, ${g_{n,n}}$ may be simplified as a real variable:
\begin{equation}
{g_{n,n}} = \max ({x_2}) = \sqrt {\frac{{{N_t} - 1}}{N_t}}.
\!\label{c-nn}
\end{equation}
Based on $\left\| {{{\bf{g}}_n}} \right\|_2^2 = 1$, Eq. (\ref{A-C-AllElements}), and Eq. (\ref{g-n1}), the rest elements are calculated by
\begin{equation}
{g_{n,q}} =  - \frac{{ - {\eta _1} + i{\kappa _1}}}{{u_{1,1}^*\sqrt {{N_t}({N_t} - 1)} }},\ q = 2, \cdots ,n - 1,n + 1, \cdots ,{N_t},
\!\label{c-nq}
\end{equation}
and
\begin{equation}
{g_{n,1}} =  - \frac{{({N_t} - 2){x_1} + {\eta _1}{x_2} + i(({N_t} - 2){y_1} - {\kappa _1}{x_2})}}{{u_{1,1}^*}},
\!\label{c-n1}
\end{equation}
where
\begin{equation}
\begin{array}{l}
{x_1} =  - \frac{{{\eta _1}}}{{\sqrt {{N_t}({N_t} - 1)} }}\\
{y_1} = \frac{{{\kappa _1}}}{{\sqrt {{N_t}({N_t} - 1)} }}.
\end{array}
\!\label{value-x1-y1}
\end{equation}
Until now, the vector ${{\bf{g}}_n}$ is calculated, and the matrix ${\bf{G}}$ is designed as
\begin{equation}
{\bf{G}} = \left[ {\begin{array}{*{20}{c}}{{\bf{\cal C}},}&{{{\bf{0}}_{{N_t} \times {N_t}(P - 1)}}} \end{array}} \right],
\!\label{Matrix-G-Result}
\end{equation}
where
\begin{equation}
{\bf{\cal C}} = {\left[ {\begin{array}{*{20}{c}}{{\bf{g}}_{1,1:{N_t}},}&{ \cdots ,}&{{{\bf{g}}_{{N_t},1:{N_t}}}} \end{array}} \right]^T},
\!\label{Matrix-C}
\end{equation}
and ${{\bf{g}}_{n,1:{N_t}}} = {[{g_{n,1}}, \cdots ,{g_{n,{N_t}}}]^T}$. 

Since the bulk energy of ${\bf{\cal C}}$ is concentrated on the diagonal elements, we select the diagonal elements to approximate ${\bf{\cal C}}$:
\begin{equation}
{\bf{\cal C}} \approx {\rm{diag}}\{ {\begin{array}{*{20}{c}}{{g_{1,1}},}&{ \cdots ,}&{g_{{N_t},{N_t}}} \end{array}}\}.
\!\label{Matrix-C-approximation}
\end{equation}
Notice that such an approximation is coincident with the asymptotic performance of ${\bf{\cal C}}$, proved in the next section. 
Then, basing on Eq. (\ref{Matrix-G}), we may generate the matrix ${{\cal {\bm B}}}$ as
\begin{equation}
{{\cal {\bm B}}} = ( {{\bf{G}} + \sum\limits_{p = 1}^P {{{\bf{K}}_p}} } ){\bf{A}}_r^{ - 1}{( {\sum\limits_{p = 1}^P {{{\bf{K}}_p}} } )^\dag },
\!\label{Matrix-F-Result}
\end{equation}
where ${\bf{G}}$ is generated according to the procedure between Eq. (\ref{optimization Problem}) and Eq. (\ref{Matrix-C-approximation}). The matrices $\sum\limits_{p = 1}^P {{{\bf{K}}_p}}$ and ${\bf{A}}_r$ are shown in Eq. (\ref{KMatrix}) and Eq. (\ref{A-Distance}), respectively, where the number of paths $P$ and distances may be estimated by OMP algorithm in Sec. \ref{sec:Parameters Estimation}.
Since $r(\sum\limits_{p = 1}^P {{{\bf{K}}_p}} ) = r({\bf{G}}) = {N_t}$ and $r({{\bf{A}}_r}) = {N_t}P$, the matrix ${{\cal {\bm B}}}$ is full-rank: $r({{\cal {\bm B}}}) = {N_t}$.
By substituting ${\bf{G}}$, $\sum\limits_{p = 1}^P {{{\bf{K}}_p}} $, and ${{\bf{A}}_r}$ into Eq. (\ref{Matrix-F-Result}), we may easily generate ${{\cal {\bm B}}}$, which is a diagonal matrix.

Due to our focus on the performance decline caused by phase fluctuations in the near-field channel, the effective proportion of ${{\cal {\bm B}}}$ is diagonal elements phases. 
Therefore, we obtain the final wavefront-transformation matrix ${{{\cal {\bm B}}}_n}$ by normalizing all elements in ${{\cal {\bm B}}}$. 

\begin{algorithm}[htb]
\renewcommand{\algorithmicrequire}{\textbf{Input:}}
\renewcommand{\algorithmicensure}{\textbf{Output:}}
\caption{The wavefront-transformation matrix design scheme.}
\label{algorithm_TransformMatrix}
\begin{algorithmic}[1]
\REQUIRE $\bm{\hat \theta} $, $\bm{\hat \phi} $, $\bm{\hat r}$, $\hat P$;
\STATE Construct a matrix ${{\bf{\hat A}}_{\theta ,\phi }}$ containing the 3-D steering vectors based on $\bm{\hat \theta} $, $\bm{\hat \phi} $, $\bm{\hat r}$, and $\hat P$;
\FOR{$n = 1:{N_t}$}
\STATE Based on Eq. (\ref{c-nn}), Eq. (\ref{c-nq}) and Eq. (\ref{c-n1}), compute the vector ${{\bf{g}}_{n,1:{N_t}}}$;
\ENDFOR
\STATE Generate matrix ${\bf{\cal C}}$ based on Eq. (\ref{Matrix-C});
\STATE Approximate ${\bf{\cal C}}$ as Eq. (\ref{Matrix-C-approximation}) by selecting the diagonal elements;
\STATE Generate matrix ${\bf{G}}$ with Eq. (\ref{Matrix-G-Result});
\STATE Calculate the matrix ${{\cal {\bm B}}}$ basing on Eq. (\ref{Matrix-F-Result});
\STATE Finally, generate the final wavefront-transformation matrix ${{{\cal {\bm B}}}_n}$ by normalizing all entries in ${{\cal {\bm B}}}$;
\ENSURE ${{{\cal {\bm B}}}_n}$
\end{algorithmic}
\end{algorithm}
The detailed design process is illustrated in Algorithm \ref{algorithm_TransformMatrix}.
Note that the designed matrix ${{{\cal {\bm B}}}_n}$ may transform the spherical wavefront to a new wavefront closer to the plane wave.
Therefore, our designed wavefront-transformation matrix can mitigate the near-field effect in the ELAA channel.

\subsection{Time-frequency-domain Projection}\label{sec:time-frequency projection matrix}
Because of the Doppler effect on the time domain and frequency domain simultaneously in the ELAA communication system, time-varying path delay causes performance degradation that the conventional DFT matrix is unable to address. 
This section aims to design a time-frequency-domain projection matrix to track the time-varying path delay. 
The key is to determine the Doppler and delay sampling intervals.
More specifically, we first determine the Doppler interval by calculating the column coherence of two response vectors related to the Doppler.
Then, to capture the Doppler and path delays of different paths, we design a matrix that contains the delay and Doppler information at a moment.
Finally, by utilizing multiple samples, we compute a time-frequency-domain matrix to track the time-varying path delay and Doppler, which contains the time-varying path delay information.

We first denote the Doppler and delay sampling intervals as $\Delta \omega$ and $\Delta \tau$.
From Eq. (\ref{Bu}) and Eq. (\ref{b-tau-Doppler}), we may calculate the column coherence between ${\bf{b}}({\tau _{p,0}},{\omega _p}t)$ and ${\bf{b}}({\tau _{q,0}},{\omega _q}t)$:
\begin{equation}
\begin{array}{l}
{\mathfrak F}({\tau _p},{\tau _q}) = \left| {{{\bf{b}}({\tau _{q,0}},{\omega _q}t)^H}{\bf{b}}({\tau _{p,0}},{\omega _p}t)} \right|\\
  \ \ \ \ \ \ \ \ \ \ \ \ = \left| \frac{{{e^{ j\pi (2{f_1} + ({N_f} - 1)){\Delta \omega }}}}}{{{N_f}}}\frac{{\sin (\pi \Delta f{N_f}{\Delta \tau })}}{{\sin (\pi \Delta f{\Delta \tau })}} \right|.
\end{array}
\!\label{ColumnCoherence-b}
\end{equation}
In order to achieve the time-frequency-domain sparsity, the column coherence ${\mathfrak F}({\tau _p},{\tau _q})$ should be as small as possible. 
Let ${\mathfrak F}({\tau _p},{\tau _q})=0$, we may obtain $\Delta \tau  = \frac{1}{{{N_f}}}$, which is coincident with the DFT matrix \cite{18Conference}.

Then, we will determine the Doppler sampling interval ${\Delta \omega }$, which can be calculated by the column coherence between two time-domain vectors.
Denote the number of samples as $N_s$ and the duration of the channel sample as $T$.
Since ${\bf{b}}({\tau _{p,0}},{\omega _p}t)$ in Eq. (\ref{b-tau-Doppler}) contains the Doppler information at a moment, we select the phase ${e^{j2\pi ((1 + \frac{{{f_{n_f}}}}{{{f_c}}}){\omega _p}t)}}$, ${n_f}=1, \cdots, {N_f}$, and construct a time-domain response vector ${{\bm{\omega}}_p}({f_{{n_f}}})$ at the ${n_f}$-th subcarrier frequency as:
\begin{equation}
\!\!\!\!{{\bm{\omega}}_p}({f_{{n_f}}}) \!=\! {\left[\!\! {\begin{array}{*{20}{c}}{{e^{j2\pi {\omega _p}(\frac{{{f_c} + {f_{{n_f}}}}}{{{f_c}}})T}},\!\!\!\!}&{ \cdots ,\!\!\!\!\!}&{{e^{j2\pi {\omega _p}(\frac{{{f_c} + {f_{{n_f}}}}}{{{f_c}}}){N_s}T}}} \end{array}} \!\!\!\right]^T}\!.
\!\label{DopplerResponseVector}
\end{equation}
The column coherence between two time-domain response vectors is calculated by
\begin{equation}
\begin{array}{l}
\!{\mathfrak F}({\omega _p},{\omega _q}) \!\!=\!\! \left| {{{\bm{\omega}}_q^H}({f_{{n_f}}}){{\bm{\omega}}_p}({f_{{n_f}}})} \right|\!\! =\!\! \left| {\frac{{\sin (\pi ({\omega _p} - {\omega _q})(\frac{{{f_c} + {f_{{n_f}}}}}{{{f_c}}}){N_s}T)}}{{\sin (\pi ({\omega _p} - {\omega _q})(\frac{{{f_c} + {f_{{n_f}}}}}{{{f_c}}})T)}}} \right|.
\end{array}
\!\label{ColumnCoherence}
\end{equation}
Similar to the procedure of calculating $\Delta \tau$, we let ${\mathfrak F}({\omega _p},{\omega _q}) = 0$ and may obtain
\begin{equation}
\begin{array}{l}
\Delta \omega  \buildrel \Delta \over = {\omega _p} - {\omega _q} = \frac{{{f_c}}}{{{N_s}T({f_c} + {f_1} + ({n_f} - 1)\Delta f)}}.
\end{array}
\!\label{ColumnCoherenceDoppler}
\end{equation}

Due to $\min({f_c}, {f_1}) \gg B$, $\Delta \omega$ is approximated as $\Delta \omega  \approx \frac{{{f_c}}}{{{N_s}T({f_c} + {f_1})}}$. Therefore, we may obtain ${\mathfrak F}({\omega _q} + \Delta \omega ,{\omega _q}) \approx 0$ at all subcarrier frequencies.

Next, since the conventional DFT matrix fails to track the Doppler effect in the frequency domain, we may design some matrices containing the Doppler and path delay. 
Additionally, the channels at each subcarrier frequency have different Doppler effect in different paths.
As a result, we design a time-frequency-domain projection matrix at time ${n_t}T$ as:
\begin{equation}
{{\cal {\bm D}}_{{n_t}}} = \left[ {\begin{array}{*{20}{c}}{{{\bf{W}}_1},}&{ \cdots ,}&{{{\bf{W}}_{{n_s}}},}&{ \cdots ,}&{{{\bf{W}}_{{N_s}}}} \end{array}} \right],
\!\label{delay-doppler-projection-matrix}
\end{equation}
where $1 \le {n_t} \le {N_s}$, $1 \le {n_s} \le {N_s}$, and ${{\bf{W}}_{{n_s}}} \in {{\mathbb{C}}^{{N_f} \times {N_f}}}$ consists of the delay-and-Doppler response vectors: 
\begin{equation}
\begin{array}{l}
{{\bf{W}}_{{n_s}}} = {\frac{1}{{\sqrt {{N_f}} }}}[{\bf{b}}(0\Delta \tau ,({n_s} - 1){n_t}\Delta \omega T), \cdots ,\\
\ \ \ \ \ \ \ \ \ \ \ \ \ \ \ \ {\bf{b}}(({N_f} - 1)\Delta \tau ,({n_s} - 1){n_t}\Delta \omega T)],
\end{array}
\!\label{W-single-time-slot}
\end{equation}
where ${\bf{b}}((n_f-1)\Delta \tau ,({n_s} - 1){n_t}\Delta \omega T)$ denotes the delay response vector at time ${n_t}T$. Specifically, $(n_f-1)\Delta \tau$ is used to capture the path delay, and $(n_s-1){\Delta \omega}$ can capture the Doppler effect of different paths.
The physical meaning of ${{\bf{W}}_{{n_s}}}$ is introducing the ${n_s}$-th Doppler sampling interval $(n_s-1){\Delta \omega}$ and $N_f$ delay sampling intervals at time ${n_t}T$ to track the delays of different paths.
If ${n_s} = 1$, the channel is static without Doppler effect.
In this case, ${{\bf{W}}_1}$ is a DFT matrix with a size of ${N_f} \times {N_f}$.
In Eq. (\ref{delay-doppler-projection-matrix}), the physical meaning of ${{\cal {\bm D}}_{{n_t}}}$ is a time-frequency-domain matrix containing $N_s$ Doppler sampling intervals and $N_f$ delay sampling intervals at time ${n_t}T$, which may track the path delays and various Doppler of different paths.

Finally, to track the time-varying path delay and Doppler, we may extend the matrix ${{\cal {\bm D}}_{{n_t}}}$ in Eq. (\ref{delay-doppler-projection-matrix}) at time ${n_t}T$ to other moments, and design the time-frequency-domain projection matrix ${{\cal {\bm D}}_d} \in {{\mathbb{C}}^{{N_f} \times {N_f}N_s^2}}$ as a block matrix:
\begin{equation}
{{\cal {\bm D}}_d} = \left[ {\begin{array}{*{20}{c}}{{{\cal {\bm D}}_1},}&{ \cdots ,}&{{{\cal {\bm D}}_{{n_t}}},}&{ \cdots ,}&{{{\cal {\bm D}}_{{N_s}}}} \end{array}} \right].
\!\label{F-d-Matrix}
\end{equation}
The physical meaning of ${{\cal {\bm D}}_d}$ is a time-frequency-domain matrix containing the Doppler information and time-varying path delay information of $N_s$ samples. 
In the mobility problem, the effect of phase shift, brought from the Doppler effect, enhances as time passes.
Our designed time-frequency-domain projection matrix ${{\cal {\bm D}}_d}$ can mitigate the phase shift effect.

For clarification, the detailed generation process is summarized in Algorithm \ref{algorithm_TimeFrequencyProjection}. 
Note that the designed matrix ${{\cal {\bm D}}_d}$ only depends on the number of time samples $N_s$, time sampling interval $T$, central carrier frequency $f_c$ and bandwidth $B$.
With matrix ${{\cal {\bm D}}_d}$, we may track the Doppler and time-varying path delay.
\begin{algorithm}[htb]
\renewcommand{\algorithmicrequire}{\textbf{Input:}}
\renewcommand{\algorithmicensure}{\textbf{Output:}}
\caption{The time-frequency-domain matrix design scheme.}
\label{algorithm_TimeFrequencyProjection}
\begin{algorithmic}[1]
\REQUIRE $N_s$, $T$, $f_c$, $B$;
\STATE Compute the column coherence ${\mathfrak F}({\tau _p},{\tau _q})$ based on Eq. (\ref{ColumnCoherence-b});
\STATE Obtain the sampling interval of delay as $\Delta \tau  = \frac{1}{{{N_f}}}$ by setting ${\mathfrak F}({\tau _p},{\tau _q}) = 0$;
\STATE According to Eq. (\ref{ColumnCoherence}), compute the column coherence ${\mathfrak F}({\omega _p},{\omega _q})$;
\STATE Let ${\mathfrak F}({\omega _p},{\omega _q}) = 0$ and determine the Doppler sampling interval as $\Delta \omega  \approx \frac{{{f_c}}}{{{N_s}T({f_c} + {f_1})}}$; 
\STATE Construct a delay-and-Doppler response matrix ${{\bf{W}}_{{n_s}}}$ as Eq. (\ref{W-single-time-slot});
\STATE Generate a time-frequency-domain projection matrix ${{\cal {\bm D}}_{{n_t}}}$ at time ${n_t}T$ as Eq. (\ref{delay-doppler-projection-matrix});
\STATE By extending ${{\cal {\bm D}}_{{n_t}}}$ to other moments, design the time-frequency-domain projection matrix ${{\cal {\bm D}}_d}$ as Eq. (\ref{F-d-Matrix});
\ENSURE ${{\cal {\bm D}}_d}$
\end{algorithmic}
\end{algorithm}

\subsection{Doppler Estimation}\label{sec:prediction method}

Basing on the wavefront-transformation matrix ${{\cal {\bm B}}_n}$ designed in Sec. \ref{sec:wavefront transform matrix}, we first mitigate the effect of phase fluctuations introduced by spherical wavefront. 
Since the channels at different subcarrier frequencies share the same distance, the wavefront-transformation matrix ${\bf{Q}} \in {{\mathbb{C}}^{{N_t}{N_f} \times {N_t}{N_f}}}$ in the frequency domain is expressed as
\begin{equation}
{\bf{Q}} = {{\bf{I}}_{{N_f}}} \otimes {{\cal {\bm B}}_n}.
\!\label{Q-Matrix}
\end{equation}
By using the time-frequency-domain projection matrix ${{\cal {\bm D}}_d}$ and two DFT matrices, i.e., ${{\bf{W}}_h} \in {{\mathbb{C}}^{{N_h} \times {N_h}}}$ and ${{\bf{W}}_v} \in {{\mathbb{C}}^{{N_v} \times {N_v}}}$,
the joint angular-time-frequency basis ${\bf{S}} \in {{\mathbb{C}}^{{N_t}{N_f} \times {N_t}{N_f}N_s^2}}$ is computed by
\begin{equation}
{\bf{S}} = {{\cal {\bm D}}_d} \otimes {{\bf{W}}_h} \otimes {{\bf{W}}_v} = {\bf{\cal U}}{\bf{\cal V}},
\!\label{S-Matrix}
\end{equation}
where ${\bf{\cal U}} = {{\bf{I}}_{{N_f}}} \otimes {{\bf{W}}_{{N_h}}} \otimes {{\bf{W}}_{{N_v}}} \in {{\mathbb{C}}^{{N_t}{N_f} \times {N_t}{N_f}}}$ is an orthogonal angular-domain basis, and ${\bf{\cal V}} =  {{{\cal {\bm D}}_d} \otimes {{\bf{I}}_{{N_h}}} \otimes {{\bf{I}}_{{N_v}}}} \in {{\mathbb{C}}^{{N_t}{N_f} \times {N_t}{N_f}N_s^2}}$ is a time-frequency-domain basis.
After mitigating the near-field effects, with the angular-time-frequency basis {\bf{S}}, the vectorized channel ${{\bf{\bar h}}_{\nu}}(t)$ in Eq. (\ref{ChannelVectorized}) is projected onto the angular-time-frequency domain:
\begin{equation}
{{\bf{{{\Omega }}}}_{\nu}}(t) = {{\bf{S}}^{\dag} }{\bf{Q}}{{\bf{\bar h}}_{\nu}}(t),
\!\label{w-vector}
\end{equation}
where ${{\bf{{{\Omega }}}}_{\nu}}(t) \in {{\mathbb{C}}^{{N_t}{N_f}N_s^2 \times 1}}$ is the vectorized channel in the angular-time-frequency domain, and ${\bf{S}}^{\dag} = {{\cal {\bm D}}_d^{\dag}} \otimes {{\bf{W}}_h^H} \otimes {{\bf{W}}_v^H}$.
Most of the entries in ${{\bf{{{\Omega }}}}_{\nu}}(t)$ may be close to zero because the number of paths is less than the size of ${{\bf{{{\Omega }}}}_{\nu}}(t)$, i.e, $P \ll {N_t}{N_f}N_s^2$. 
Define a positive threshold ${\gamma _1}$ that is close to 1. The number of non-negligible entries is determined by
\begin{equation}
{\cal N} = \mathop {\arg \min }\limits_{\cal N} \sum\limits_{n = 1}^{\cal N} {\left| {{\chi _{n}}{\rm{(}}t{\rm{)}}} \right|}  \ge {\gamma _1}\sum\limits_{n = 1}^{{N_t}{N_f}N_s^2} {\left| {{\chi _{n}}{\rm{(}}t{\rm{)}}} \right|},
\!\label{N-determine}
\end{equation}
where ${{\chi _{n}}{\rm{(}}t{\rm{)}}}$ is the $n$-th entry of ${{\bf{{{\Omega }}}}_{\nu}}(t)$ and is expressed as
\begin{equation}
{\chi _{n}}{\rm{(}}t{\rm{)}} = \sum\limits_{{m} = 1}^{{N_t}{N_f}} \sum\limits_{{n_f} = 1}^{{N_f}} {} {\chi _{n,m}}{\rm{(}}t,{n_f}{\rm{)}},
\!\label{w-n-determine}
\end{equation}
and
\begin{equation}
{\chi _{n,m}}{\rm{(}}t,{n_f}{\rm{) = }}\sum\limits_{p = 1}^P {} {\beta _p}{\alpha _{m,p}}{g_{{n, m}}}{e^{j2\pi (\frac{{{f_{{n_f}}} + {f_c}}}{{{f_c}}}){\omega _p}t}}.
\!\label{w-n-nf}
\end{equation}
The projection of channel in the angular domain ${\alpha _{m,p}}$ is time-invariant and generated by the ${{\bf{\cal U}}}$.
Also, ${g_{{n, m}}}$ is the $m$-th entry of the $n$-th non-negligible row vector in ${{\bf{\cal V}}^\dag }$. 
The vectorized channel is approximated by
\begin{equation}
{{\bf{\bar h}}_{\nu}}(t) \approx \sum\limits_{n = 1}^{\cal N} {} {\chi _{n}}{\rm{(}}t{\rm{)}}{{\bf{Q}}^{ - 1}}{{\bf{s}}_n},
\!\label{channel-approximate}
\end{equation}
where ${{\bf{s}}_n}$ is the $n$-th column vector of ${\bf{S}}$.
Next, we adopt the MP method to estimate Doppler. 
For notational simplicity, we rewrite ${\chi _{n,m}}(t,{n_f})$ as ${\chi _{n,m}}(t)$.
Define an MP matrix ${{\bf{D}}_1} \in {{\mathbb{C}}^{Q \times ({N_s} - Q + 1)}}$ at the $n_f$-th subcarrier frequency as
\begin{equation}
\begin{array}{l}
\!\!\!\!\!\!\!{{\bf{D}}_1} \!=\! \left[ \begin{array}{l}
{\chi _{n,m}}{\rm{(}}T{\rm{)}}, \ \ \cdots ,\ \ {\chi _{n,m}}{\rm{(}}({N_s} - Q + 1)T{\rm{)}}\!\!\!\!\\
\ \ \ \ \ \vdots \ \ \ \ \ \ \ \ \ddots \ \ \ \ \ \ \ \ \ \ \ \ \ \ \vdots \\
{\chi _{n,m}}{\rm{(}}QT{\rm{)}}, \cdots ,\ \ \ \ \ \ {\chi _{n,m}}{\rm{(}}{N_s}T{\rm{)}}
\end{array} \ \ \right]\\
 \ = {{\bf{E}}_{1}}{\bf{Z}}{\bf{R}}{{\bf{E}}_{2}},
\end{array}\!\!\!
\!\!\!\!\!\!\label{MP matrix-D1}
\end{equation}
where the pencil size $Q$ satisfies $P < Q < {N_s} - P + 1$, ${\bf{R}} = {\rm{diag}}\left\{ {{\beta _1}{\alpha _{m,1}}{g_{{n, m}}}, \cdots ,{\beta _P}{\alpha _{m,P}}{g_{{n, m}}}} \right\}$, and
\begin{equation}
\!\!{{\bf{E}}_{1}} \!\!=\!\! \left[\!\! \begin{array}{l}
\ \ \ \ \ \ \ \ \ 1,\ \ \ \ \ \ \ \ \ \ \ \ \ \cdots, \ \ \ \ \ \ \ \ \ \ 1\\
\ \ \ \ \ \ \ \ \ \ \vdots \ \ \ \ \ \ \ \ \ \ \ \ \ \ \ddots \ \ \ \ \ \ \ \ \ \ \ \ \vdots \\
{e^{j2\pi \frac{{{f_c} + {f_{{n_f}}}}}{{{f_c}}}{\omega _1}(Q-1)T}}, \cdots ,{e^{j2\pi \frac{{{f_c} + {f_{{n_f}}}}}{{{f_c}}}{\omega _P}(Q-1)T}}
\end{array} \!\!\right]\!\!,
\!\label{MP matrix-E-left}
\end{equation}
\begin{equation}
{{\bf{E}}_{2}} = \left[ \begin{array}{l}
\ 1,\ \ \ \ \ \ \cdots , \ \ \ \ \ \ {e^{j2\pi \frac{{{f_c} + {f_{{n_f}}}}}{{{f_c}}}{\omega _1}({N_s} - Q)T}}\\
\ \ \vdots \ \ \ \ \ \ \ \ddots \ \ \ \ \ \ \ \ \ \ \ \ \ \ \ \ \vdots \\
\ 1,\ \ \ \ \ \ \cdots ,\ \ \ \ \ \ {e^{j2\pi \frac{{{f_c} + {f_{{n_f}}}}}{{{f_c}}}{\omega _P}({N_s} - Q)T}}
\end{array} \right],
\!\label{MP matrix-E-right}
\end{equation}
\begin{equation}
{\bf{Z}} = {\rm{diag}}\!\left\{ {\begin{array}{*{20}{c}}{{e^{j2\pi \frac{{{f_c} + {f_{{n_f}}}}}{{{f_c}}}{\omega _1}T}},}&{ \cdots ,}&{{e^{j2\pi \frac{{{f_c} + {f_{{n_f}}}}}{{{f_c}}}{\omega _P}T}}} \end{array}}\!\right\}.
\!\!\!\label{MP matrix-Z}
\end{equation}
Select the first $({N_s} - Q)$ and the last $({N_s} - Q)$ columns of ${{\bf{D}}_1}$ as ${{\bf{D}}_{1,1:({N_s} - Q)}}$ and ${{\bf{D}}_{1,2:({N_s} - Q + 1)}}$, respectively. 
The matrix ${\bf{Z}}$ is estimated by
\begin{equation}
{\bf{\hat Z}} = eig( {{{\bf{D}}_{1,2:({N_s} - Q + 1)}}{\bf{D}}_{1,1:({N_s} - Q)}^\dag}).
\!\label{Z matrix-estimation}
\end{equation}
The Doppler of the $p$-th path is estimated as 
\begin{equation}
{\hat{\omega} _p} = \frac{angle({\hat{z}_p}){f_c}}{2\pi({{f_c} + {f_{{n_f}}}})T},
\!\label{Doppler-estimation}
\end{equation}
where ${\hat{z}_p}$ is the $p$-th entry of ${\bf{\hat Z}}$.
According to Eq. (\ref{MP matrix-E-left}) and Eq. (\ref{MP matrix-E-right}), we may easily obtain the estimations of ${{\bf{E}}_{1}}$ and ${{\bf{E}}_{2}}$ as ${{\bf{\hat E}}_{1}}$ and ${{\bf{\hat E}}_{2}}$, respectively.
From Eq. (\ref{MP matrix-D1}), we also estimate ${\bf{R}}$ as ${\bf{\hat R}}$.
Define a new MP matrix ${{\bf{D}}_2} \in {{\mathbb{C}}^{Q \times ({N_s} - Q + 1)}}$ as
\begin{equation}
\!\!\!\!{{\bf{D}}_2} \!\!=\!\! \left[ \begin{array}{l}
\ {\chi _{n,m}}{\rm{(2}}T{\rm{)}}, \ \ \ \ \ \cdots , \ {\chi _{n,m}}(({N_s} - Q + 2)T)\\
\ \ \ \ \ \ \ \ \vdots \ \ \ \ \ \ \ \ \ \ \ \ \ddots  \ \ \ \ \ \ \ \ \ \ \ \ \ \ \vdots \\
\!\!{\chi _{n,m}}((Q + 1)T), \ \cdots ,\ \ \ {\chi _{n,m}}(({N_s} + 1)T)
\end{array} \!\!\right],
\!\label{MP matrix-D2}
\end{equation}
which is estimated by 
\begin{equation}
{{\bf{\hat D}}_2} = {{\bf{\hat E}}_{1}}{\bf{{{\hat Z}}}}{\bf{\hat E}}_{1}^\dag {{\bf{D}}_1}{\bf{\hat E}}_{2}^\dag {{\bf{\hat E}}_{2}}.
\!\label{MP matrix-D2estimation}
\end{equation}
By selecting the last entry from ${{\bf{\hat D}}_2}$, we may estimate ${\hat \chi _{n,m}}{\rm{((}}{N_s} + 1)T{\rm{)}}$. 
Denote the number of predicted samples as ${N_d}$.
Update Eq. (\ref{MP matrix-D2}) by removing the first column and appending a new column at last based on the last $Q$ predictions. 
Then, repeat Eq. (\ref{MP matrix-D2estimation}) $({N_d} - 1)$ times by replacing ${{\bf{D}}_1}$ with ${{\bf{\hat D}}_2}$. 
We may predict ${\hat \chi _{n,m}}{\rm{((}}{N_s} + {N_d})T{\rm{)}}$, which is a simplified notation of ${\hat \chi _{n,m}}{\rm{((}}{N_s} + {N_d})T, {n_f}{\rm{)}}$ in Eq. (\ref{w-n-nf}).
Furthermore, predict ${\hat \chi _{n,m}}{\rm{((}}{N_s} + {N_d})T, {n_f}{\rm{)}}$ at each subcarrier frequency by repeating the prediction process of ${\hat \chi _{n,m}}{\rm{((}}{N_s} + {N_d})T, {n_f}{\rm{)}}$ between Eq. (\ref{MP matrix-D1}) and Eq. (\ref{MP matrix-D2estimation}) $({N_f} -1)$ times. 
We may predict ${\hat \chi _{n}}({\rm{(}}{N_s} + {N_d})T)$ as:
\begin{equation}
\begin{array}{l}
\!{\hat \chi _{n}}({\rm{(}}{N_s} + {N_d})T) \!=\!\!\!\!\! \sum\limits_{{m} = 1}^{{N_t}{N_f}}\!\! \sum\limits_{{n_f} = 1}^{{N_f}} {}\!{\hat \chi _{n,m}}{\rm{(}}{\rm{(}}{N_s} + {N_d})T,{n_f}{\rm{)}}.
\end{array}
\!\!\!\!\label{w-n-determine-estimation}
\end{equation}
The vectorized channel at time $t$ ($({N_s} + 1)T \le t \le ({N_s} + {N_d})T$) is predicted as
\begin{equation}
{{\bf{\hat h}}_{\nu}}{\rm{(}}t{\rm{)}} \approx \sum\limits_{n = 1}^{\cal N} {\hat \chi _n}{\rm{(}}t{\rm{)}}{{\bf{Q}}^{ - 1}}{{\bf{s}}_n}.
\!\label{channel prediction}
\end{equation}
The details of our proposed WTMP channel prediction method are summarized in Algorithm \ref{algorithm_WTMP}.
Notice that $\bm{\hat \theta} $, $\bm{\hat \phi} $ and $\bm{\hat r}$ in step 1 may also be estimated by some super-resolution methods, e.g., MUSIC and ESPRIT.
However, the super-resolution methods may introduce enormous computational complexity due to multi-dimensional search.
Compared to the super-resolution methods, our adopted OMP algorithm in step 1 needs less computational complexity.
By increasing the sampling grid points of EOD, AOD, and distance in step 1, the estimation accuracy of angles and distance may increase.

\begin{algorithm}[htb]
\renewcommand{\algorithmicrequire}{\textbf{Input:}}
\renewcommand{\algorithmicensure}{\textbf{Output:}}
\caption{The proposed WTMP channel prediction scheme.}
\label{algorithm_WTMP}
\begin{algorithmic}[1]
\REQUIRE ${{\bf{H}}_{\nu}}(t)$, ${{\bf{\bar A}}_u}$, $N_d$
\STATE Estimate $\bm{\hat \theta} $, $\bm{\hat \phi} $ and $\bm{\hat r}$ via OMP algorithm;
\STATE Design the wavefront-transformation matrix ${{\cal {\bm B}}_n}$ with Algorithm \ref{algorithm_TransformMatrix};
\STATE Design the time-frequency-domain projection matrix ${{\cal {\bm D}}_d}$ with Algorithm \ref{algorithm_TimeFrequencyProjection};
\STATE Transform the channel onto angular domain and time-frequency domain with Eq. (\ref{w-vector});
\STATE Select non-negligible entries of ${{\bf{{{\Omega }}}}_{\nu}}(t)$ with Eq. (\ref{N-determine});
\STATE Basing on Eq. (\ref{MP matrix-D1}), Eq. (\ref{Z matrix-estimation}) and Eq. (\ref{Doppler-estimation}), estimate the Doppler ${\hat{\omega} _p}$. 
\FOR{${n_d} = 1:{N_d}$}
\FOR{${n_f} = 1:{N_f}$}
\STATE Compute the MP matrix ${{\bf{\hat D}}_2}$ by Eq. (\ref{MP matrix-D2estimation});
\STATE Select the last entry from ${{\bf{\hat D}}_2}$ as the prediction of ${\hat \chi _{n,m}}{\rm{((}}({N_s} + {n_d})T),{n_f}{\rm{)}}$;
\ENDFOR
\STATE Predict ${\hat \chi _{n}}({\rm{(}}{N_s} + {n_d})T)$ as in Eq. (\ref{w-n-determine-estimation});
\ENDFOR
\STATE Reconstruct and predict channel as ${{\bf{\hat h}}_{\nu}}{\rm{(}}t+ {N_d}T{\rm{)}}$ in Eq. (\ref{channel prediction}) by updating ${\chi _{n}}(t)$ as ${\hat \chi _{n}}(t + {N_d}T)$;
\ENSURE ${{\bf{\hat h}}_{\nu}}{\rm{(}}t+ {N_d}T{\rm{)}}$
\end{algorithmic}
\end{algorithm}

In Algorithm 3, the computation complexity is dominated by step 1, step 9, and step 14.
In the step 1, the OMP algorithm needs $\hat P$ iterations, and the computation complexity of the $p$-th iteration is $\mathcal{O}({M_\theta }{M_\phi }{M_r}{N_t}) + \mathcal{O}({N_t}{p^2}) + \mathcal{O}({p^{2.37}})$.
Step 9 has a complexity order of $\mathcal{O}({Q^2}({N_s} - Q + 1))$.
Repeating step 9 ${\cal N}{N_d}{N_t}{N_f^2}$ times, step 14 has a complexity order of $\mathcal{O}({\cal N}{N_d}{N_t}{N_f}^2{Q^2}({N_s} - Q + 1)) + \mathcal{O}({\cal N}{N_t^{2.37}}{N_f^{2.37}})$.
The global complexity of the WTMP method is $\mathcal{O}({\hat P}{M_\theta }{M_\phi }{M_r}{N_t}) + \mathcal{O}({\cal N}{N_d}{N_t}{N_f^2}{Q^2}({N_s} - Q + 1)) + \mathcal{O}({\cal N}{N_t^{2.37}}{N_f^{2.37}})$.

\section{Performance Analysis of The WTMP Prediction Method}\label{sec:performance analysis}

In this section, we start the performance analysis of our proposed WTMP prediction method by proving the asymptotic performance of the designed matrix ${{\cal {\bm B}}_n}$.
Then, the asymptotic prediction error of the WTMP method is derived, for the case of enough number of samples are available and the BS has finite antennas. 
Finally, we derive the asymptotic prediction error under the condition with the enough BS antennas and finite samples.
More details will be shown below.

\begin{Proposition}\label{proposition1} 
If the number of the BS antennas is large enough, the designed wavefront-transformation matrix ${{\cal {\bm B}}_n}$ is determined by the angles and distance between the BS antenna array and the scatterer or the UE.

\end{Proposition}
\begin{proof} 
According to Eq. (\ref{Matrix-F-Result}), the wavefront-transformation matrix ${{\cal {\bm B}}_n}$ is determined by ${\bf{G}}$, $\sum\limits_{p = 1}^P {{{\bf{K}}_p}}$ and ${\bf{A}}_r$, where $\sum\limits_{p = 1}^P {{{\bf{K}}_p}}$ is independent of the angles, and ${\bf{A}}_r$ is related to angles and distance.
Next, we aim to prove that the matrix ${\bf{G}}$ is asymptotically independent of angles and distance.
Thus, we transform the proof to a sub-problem:
\begin{equation}
\mathop {\lim }\limits_{{N_t} \to \infty } {\bf{G}} = \left[ {\begin{array}{*{20}{c}}{{{\bf{I}}_{{N_t}}},}&{{{\bf{0}}_{{N_t} \times {N_t}(P - 1)}}} \end{array}} \right].
\!\label{G asymptotic}
\end{equation}
According to Eq. (\ref{c-nn}), Eq. (\ref{c-nq}), and Eq. (\ref{c-n1}), with the number of antennas increasing, the entries of the $n$-th row vector of ${\cal C}$ in Eq. (\ref{Matrix-C}) are calculated by
\begin{equation}
\mathop {\lim }\limits_{{N_t} \to \infty } {g_{n,n}} = \mathop {\lim }\limits_{{N_t} \to \infty }\sqrt {\frac{{{N_t} - 1}}{N_t}} = 1,\ n = 1, \cdots ,{N_t},
\!\label{c-nn asymptotic}
\end{equation}
\begin{equation}
\begin{array}{l}
\mathop {\lim }\limits_{{N_t} \to \infty } \left\| {{g_{n,q}}} \right\|_2^2 \!=\!\! \mathop {\lim }\limits_{{N_t} \to \infty } \frac{{{\eta _1}^2 + {\kappa _1}^2}}{{\left\| {u _{1,1}^*} \right\|_2^2{N_t}({N_t} - 1)}}
\!=\!\! \mathop {\lim }\limits_{{N_t} \to \infty } \frac{1}{{{N_t}({N_t} - 1)}} 
\\
\ \ \ \ \ \ \ \ \ \ \ \ \ \ \ \ \ = 0,\ q = 2, \cdots ,n - 1,n + 1, \cdots ,{N_t},
\end{array}
\!\!\!\!\!\!\!\label{c-nq asymptotic}
\end{equation}
and
\begin{equation}
\begin{array}{l}
\!\mathop {\lim }\limits_{{N_t} \to \infty } \left\| {{g_{n,1}}} \right\|_2^2 = \mathop {\lim }\limits_{{N_t} \to \infty } 1 \!-\! \left\| {{g_{n,n}}} \right\|_2^2 - \!\!\sum\limits_{q = 2,q \ne n}^{{N_t}} {\left\| {{g_{n,q}}} \right\|_2^2} \\
 \ \ \ \ \ \ \ \ \ \ \ \ \ \ \ \ \ = \mathop {\lim }\limits_{{N_t} \to \infty } \frac{1}{{{N_t}({N_t} - 1)}} = 0.
\end{array}
\!\label{c-n1 asymptotic}
\end{equation}
Therefore, 
\begin{equation}
\mathop {\lim }\limits_{{N_t} \to \infty } {\cal C} = {{\bf{I}}_{{N_t}}}.
\!\label{C asymptotic}
\end{equation}
In other words,
\begin{equation}
\mathop {\lim }\limits_{{N_t} \to \infty } {\bf{G}} = \left[ {\begin{array}{*{20}{c}}{{{\bf{I}}_{{N_t}}},}&{{{\bf{0}}_{{N_t} \times {N_t}(P - 1)}}} \end{array}} \right].
\!\label{G asymptotic proof}
\end{equation}
Thus, Proposition 1 is proved.
\end{proof}
Remarks: 
The energy of the designed matrix ${\cal C}$ is concentrated on the diagonal entries. 
When the number of the BS antenna elements is large enough, we may capture nearly all energy of matrix ${\cal C}$, provided that we select the diagonal elements to approximate ${\cal C}$.
As a result, this Proposition is in line with the approximation of ${\cal C}$ in Sec. \ref{sec:wavefront transform matrix}.
Proposition 1 is also the prior basis of the following performance analysis.

Denote the vectorized form of the observation sample at time $t$ by ${{{{\bf{\tilde h}}}_{\nu}}(t)}$: ${{\bf{\tilde h}}_{\nu}}(t) = {{\bf{\bar h}}_{\nu}}(t) + {\bf{n}}(t)$,
where ${\bf{n}}(t)$ is the temporally independent identically distributed (i.i.d.) Gaussian white noise with zero mean and element-wise variance.
Considering an ELAA with a finite number of BS antennas, if the number of channel samples is large enough, the performance of our proposed WTMP method will be analyzed in Proposition 2.

\begin{Proposition}\label{proposition2} 
For an arbitrary CSI delay ${t_\tau }$, the asymptotic prediction error of the WTMP method yields:
\begin{equation}
\mathop {\lim }\limits_{{N_s} \to \infty } \frac{{\left\| {{{\bf{{\hat h}}}_{\nu}}(t+{t_\tau }) - {{\bf{\bar h}}_{\nu}}(t+{t_\tau })} \right\|_2^2}}{{\left\| {{{\bf{\bar h}}_{\nu}}(t+{t_\tau })} \right\|_2^2}} = 0,
\!\label{PredictionError asymptotic}
\end{equation}
providing that the pencil size satisfies $P < Q < {N_s} - P + 1$.
\end{Proposition}
\begin{proof} 
This Proposition is a generalization of Theorem 1 in \cite{MDMP} when the noise is temporal i.i.d., and the number of samples is large enough. 
According to Eq. (\ref{MP matrix-D1}), denote an MP matrix generated by $N_s$ observation samples as ${{\bf{\tilde D}}_1} = {{\bf{D}}_1} + {\bf{N}}$, where ${\bf{N}}$ is a noise matrix.
We may prove the Proposition as follows: Firstly, compute the correlation matrix of ${{\bf{\tilde D}}_1}$: 
${{\bf{\tilde R}}} = \mathop {\lim }\limits_{{N_s} \to \infty } E\left\{ {{{\bf{D}}_1}{\bf{D}}_1^H} \right\} + {\sigma ^2}{{\bf{I}}_{{Q}}}$, where the expectation is taken over time. 
Then, perform the SVD of ${{\bf{\tilde R}}}$ and estimate the Doppler. One may easily obtain that $\mathop {\lim }\limits_{{N_s} \to \infty } {\hat \omega _p} = {\omega _p}, p=1,\cdots, P$ and the prediction error converges to zero. 
The detailed proof is omitted.
\end{proof} 
Remarks: 
Given enough samples, Proposition \ref{proposition2} indicates that the channel prediction error converges to zero when the noise is i.i.d.. 
 
However, Proposition \ref{proposition2} requires too many samples and disregards the fact that the ELAA deploys a large number of BS antennas.
In the following, we will break these constraints and derive the asymptotic performance with enough BS antennas.
Before the analysis, we introduce a technical assumption.

\begin{assumption}\label{Assumption2}
The normalized relative error of the transform matrix ${{\cal {\bm B}}_n}$ yields:
\begin{equation}
\mathop {\lim }\limits_{{N_t} \to \infty } \frac{{\left\| {{{\cal {\bm B}}_n}{{\bf{A}}_{\nu}} - {{\bf{A}}_{{\rm{plane}}}}} \right\|_F^2}}{{{N_t}}} = {\varepsilon ^2}.
\!\label{NMSE F}
\end{equation}
\end{assumption}
Remarks: 
The sizes of ${{\bf{A}}_{\nu}}$ and ${{\bf{A}}_{{\rm{plane}}}}$ are ${N_t \times P}$.
In an arbitrary path, ${{\cal {\bm B}}_n}$ transforms one column vector of ${{\bf{A}}_{\nu}}$ and the normalized relative error ought to be finite.
Furthermore, due to the limited number of paths, the normalized relative error should be finite when ${{\cal {\bm B}}_n}$ transforms ${{\bf{A}}_{\nu}}$.
Therefore, the assumption is generally valid.

Before the following derivation, if ${N_f}=1$, we denote the vectorized form of a narrowband far-field channel as ${{\bf{{\bar h}}}_{\rm{plane}}}(t){\rm{ = }}\sum\limits_{p = 1}^P {{c_{u,p}}} {\bf{b}}({\tau _{p,0}},{\omega _p}t) \otimes {{\bf{a}}}({\theta _p},{\phi _p})$. 
After being transformed by matrix ${{\cal {\bm B}}_n}$, the narrowband near-field channel may be asymptotically quantified as
\begin{equation}
\mathop {\lim }\limits_{{N_t} \to \infty }{{\cal {\bm B}}_n}{{\bf{{\bar h}}}_{\nu}}(t) = {{\bf{{\bar h}}}_{\rm{plane}}}(t) + \sqrt {{N_t}} {\bf{x}},
\!\label{FhFv channel}
\end{equation}
where ${\bf{x}} \in {{\mathbb{C}}^{{N_t} \times 1}}$ is an error vector, and ${{\bf{x}}^H}{\bf{x}} = {\varepsilon ^2}$. 
The vector ${\bf{x}}$ is time-invariant and may not affect the estimation accuracy of Doppler.

Based on Eq. (\ref{FhFv channel}), the asymptotic performance of our proposed WTMP method will be derived in Theorem \ref{theorem}.

\begin{theorem}\label{theorem}
Under Assumption \ref{Assumption2}, for a narrowband channel, if the number of the BS antennas is large enough, and the pencil size satisfies $P < Q < N_s - P$, the asymptotic performance of our WTMP prediction method yields: 
\begin{equation}
\mathop {\lim }\limits_{{N_t} \to \infty }\!\!\!\!\!\! \frac{{\left\| {{{{{\bf{\hat h}}}_{\nu}}(t + {t_\tau })} \!-\! {{{\bf{\bar h}}_{\nu}}(t + {t_\tau })}} \right\|_2^2}}{{\left\| {{{{\bf{\bar h}}_{\nu}}(t + {t_\tau })}} \right\|_2^2}} = 0,
\!\!\!\label{theorem1}
\end{equation}
providing that $2P+2$ samples are accurate enough, i.e.,
\begin{equation}
\mathop {\lim }\limits_{{N_t} \to \infty } \frac{{\left\| {{{{\bf{\tilde h}}}_{\nu}}(t) - {{{\bf{\bar h}}}_{\nu}}(t)} \right\|_2^2}}{{\left\| {{{{\bf{\bar h}}}_{\nu}}(t)} \right\|_2^2}} = 0.
\!\!\!\label{accurate samples}
\end{equation}
\end{theorem}
\begin{proof} 
The detailed proof can be found in Appendix \ref{appendix:Theorem 1}.
\end{proof} 
Remarks:
The assumption in Eq. (\ref{accurate samples}) is a mild technology assumption, which can be fulfilled by some non-linear signal processing technologies even in the case of pilot contamination existing in the multi-user multi-cell scenario \cite{Yin16TSP}. 
Compared to Proposition \ref{proposition2}, with the help of more BS antennas, we obtain a better result that only finite samples are needed to achieve asymptotically error-free performance.

\section{Numerical results}\label{sec:simulation}
In this section, we first describe the simulation channel model and then provide numerical results to show the performance of our proposed scheme. 
Basing on the clustered delay line (CDL) channel model of 3GPP, we add an extra distance parameter to generate the simulation channel model.
The channel model consists of 9 scattering clusters, 20 rays in each cluster, and 180 propagation paths.
The extra distance parameter is the distance from the BS antenna array to the scatterers or the UE, which is modelled as a random variable uniformly distributed during the interval $[30\ {\rm{m}},45\ {\rm{m}}]$.
The root mean square (RMS) angular spreads of AOD, EOD, AOA and EOA are $71.9^\circ $, $54.4^\circ $, $120.5^\circ $ and $38.9^\circ $. 
The detailed simulation parameters are listed in Table 1.
We consider a 3D Urban Macro (3D UMa) scenario, where the UEs move at 60 km/h and 120 km/h.
The carrier frequency is 39 GHz, and the bandwidth is 20 MHz with 30 kHz subcarrier spacing.
One slot contains 14 OFDM symbols and has a duration of 0.5 ms. 
Each UE sends one sequence of Sounding Reference Signal (SRS) in a time slot.
One channel sample is available for each slot.
The antenna configuration is $(\underline{M},\underline{N}, \underline{P})$, where $\underline{M}$ is the number of horizontal antenna elements, $\underline{N}$ is the number of vertical antenna elements, and $\underline{P}$ denotes the number of polarization. 
The horizontal and vertical antenna spacings are both $0.5\lambda $. 
The BS antenna is equipped with a $2\times 256$ UPA. 
Based on Eq. (\ref{PhaseDiscrepancyNearfieldDistance}) and Eq. (\ref{PhaseDiscrepancyFarfieldDistance}), the approximation region is $[6.90\ {\rm{m}},252\ {\rm{m}}]$.
In the OMP algorithm, the numbers of sampling grid points for EOD, AOD and distance are 30, 900 and 360, respectively. 
The DL precoder is eigen-based zero-forcing (EZF) \cite{EZF}.
To assess the prediction method performance, we introduce three metrics, i.e., the DL SE, the DL prediction error, and the normalized mean square error (NMSE) of the near-field channel ${{\bf{H}}_{\nu}}$ after being transformed by matrix ${{\cal {\bm B}}_n}$.

\begin{table}[!bht]
\caption{The main simulation parameters.}
\label{table I}
\centering
\begin{tabular}{|l|p{12em}|}
\hline
Scenario & 3D Urban Macro (3D UMa)\\
\hline
Carrier frequency (GHz) & 39\\
\hline
Bandwidth (MHz) & 20 \\
\hline
Subcarrier spacing (kHz) & 30\\
\hline
Number of UEs & 16\\
\hline
BS antenna configuration & $(\underline{M},\underline{N}, \underline{P}) = (2,256,1)$, $(1,256,1)$, $(16,32,1)$, $(d_h^{{\rm{tx}}},d_v^{{\rm{tx}}})=(0.5\lambda,0.5\lambda)$\\
\hline
UE antenna configuration & $(\underline{M},\underline{N}, \underline{P}) = (1,1,2)$, the polarization angles are $0^\circ $ and $90^\circ $\\
\hline
CSI delay (ms) & 16\\
\hline
UEs speed (km/h) & 60, 120, 150\\
\hline
\end{tabular}
\end{table}

\begin{figure}[!htb]
\centering
\includegraphics[width=3.3in]{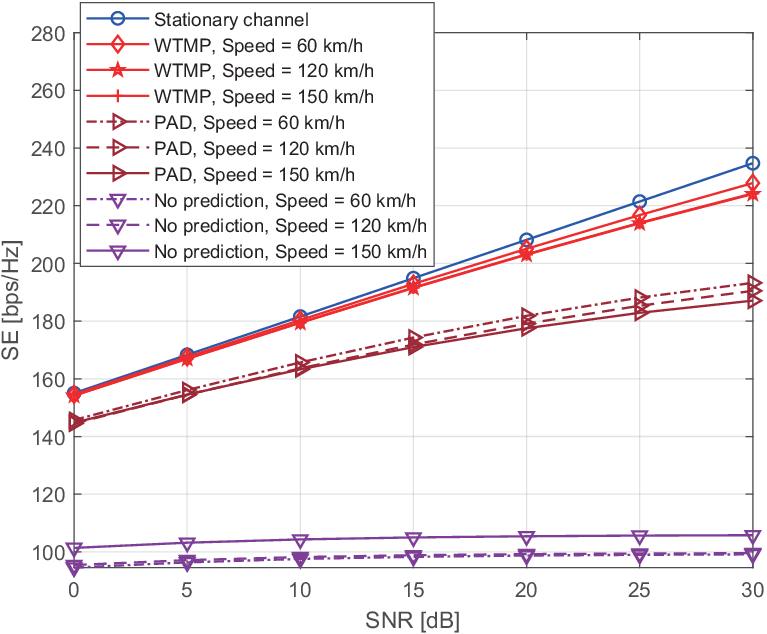}
\caption{The SE versus SNR, the BS has 512 antennas.}
\label{fig:2}
\end{figure}
Fig. \ref{fig:2} depicts the performance of different prediction methods when the UEs move at 60 km/h, 120 km/h and 150 km/h.
The CSI delay is relatively large, i.e., 16 ms.
The DL SE is calculated by $\sum\limits_{{\overline{u}} = 1}^{N_\text{UE}} {E\left\{ {{{\log }_2}(1 + {\rm{SINR}}_{\overline{u}})} \right\}} $ averaged over time and frequency, where ${\rm{SINR}}_{\overline{u}}$ is the signal-to-noise ratio of the ${\overline{u}}$-th UE and ${N_\text{UE}}$ is the number of UEs.
The ideal setting is referred as ``Stationary channel", where the DL SE achieves an upper bound of performance.
The curves labelled as ``No prediction" are the results without channel prediction.
We select the PAD channel prediction method in \cite{Yin20JSAC} as reference curves.
We may observe that the PAD method only achieves moderate prediction gains, given that the path delays are time-varying and the wavefront is spherical.
It may also be observed that our proposed method approaches the ideal setting even at a speed of 150 km/h and a CSI delay of 16 ms.
It is because our proposed method may effectively address the effects brought by the time-varying path delay and near-field radiation.

\begin{figure}[!htb]
\centering
\includegraphics[width=3.3in]{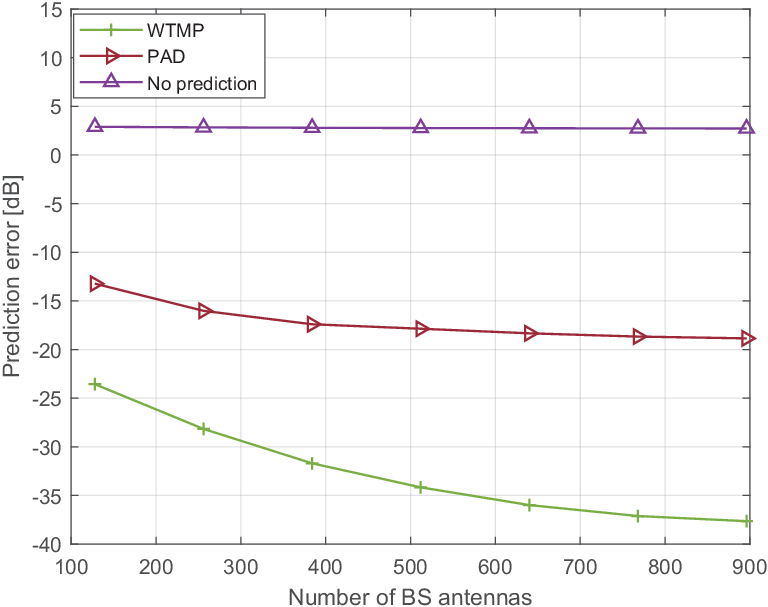}
\caption{The prediction error versus the number of BS antennas, the UEs move at 120 km/h.}
\label{fig:3}
\end{figure}
\begin{figure}[!htb]
\centering
\includegraphics[width=3.3in]{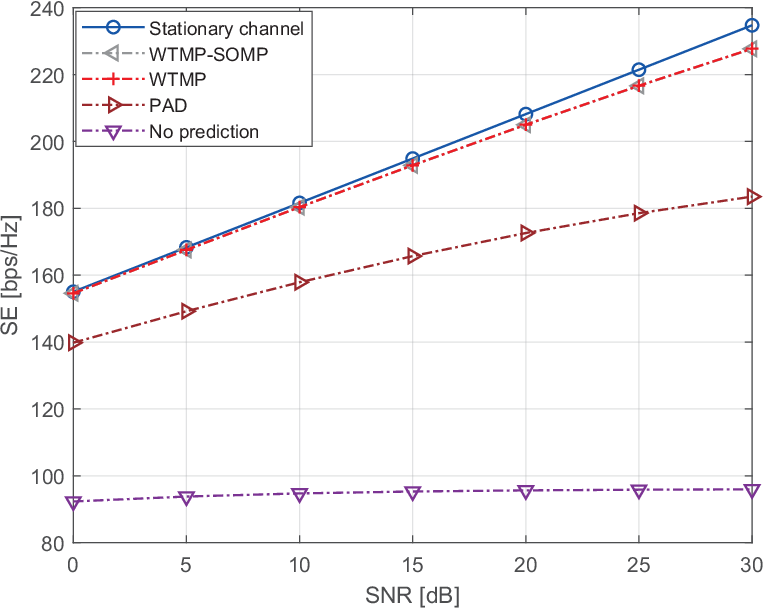}
\caption{The SE versus SNR, the BS is equipped with 512 antennas, multiple velocity levels of UEs, i.e., four at 30 km/h, four at 60 km/h, four at 90 km/h and four at 120 km/h.}
\label{fig:4}
\end{figure}
Fig. \ref{fig:3} compares the prediction errors of different methods as the number of BS antennas increases. 
The DL prediction error is computed as $10\log \left\{ {E\left\{ {\frac{{\left\| {{{{\bf{\hat H}}}_{\nu}} - {{\bf{H}}_{\nu}}} \right\|_F^2}}{{\left\| {{{\bf{H}}_{\nu}}} \right\|_F^2}}} \right\}} \right\}$, which is averaged over time, frequency and UEs.
Our proposed WTMP method outperforms the PAD method, and the prediction error asymptotically converges to zero.
It is also in line with Theorem \ref{theorem}.

Fig. \ref{fig:4} gives the SEs of different prediction methods as multiple UEs move at different velocities, i.e., every four UEs at 30 km/h, 60 km/h, 90 km/h and 120 km/h, respectively.
The curve labelled as ``WTMP-SOMP" is the result of the SOMP algorithm to estimate the distance and angles.
We may also observe that our proposed method still outperforms the PAD method and is close to the upper bound of SE.

\begin{figure}[!htb]
\centering
\includegraphics[width=3.3in]{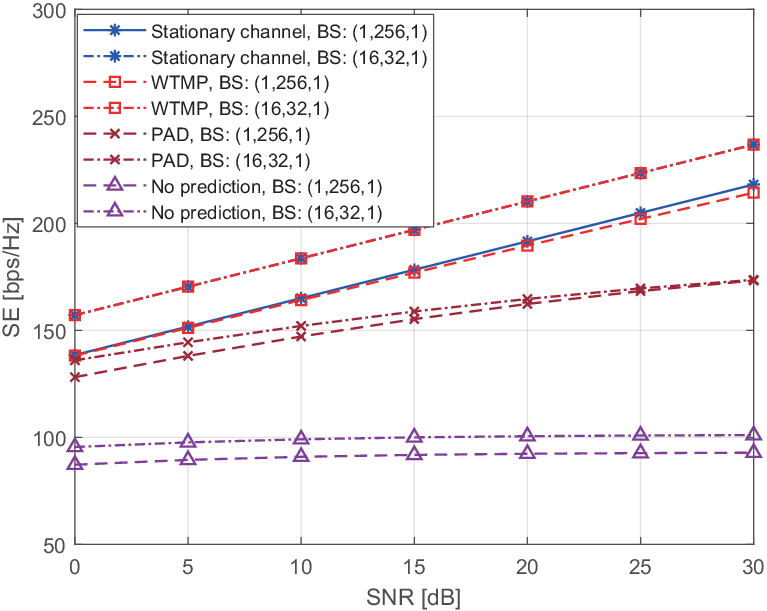}
\caption{The SE versus SNR, the UEs move at 120 km/h.}
\label{fig:5}
\end{figure}
In Fig. \ref{fig:5}, we show the SEs of different prediction methods when the BS is equipped with different antenna arrays, e.g., $(1,256,1)$ and $(16,32,1)$.
We also observe that our proposed method still outperforms the PAD method when the BS antenna configuration is a UPA or a ULA.

\begin{figure}[!htb]
\centering
\includegraphics[width=3.3in]{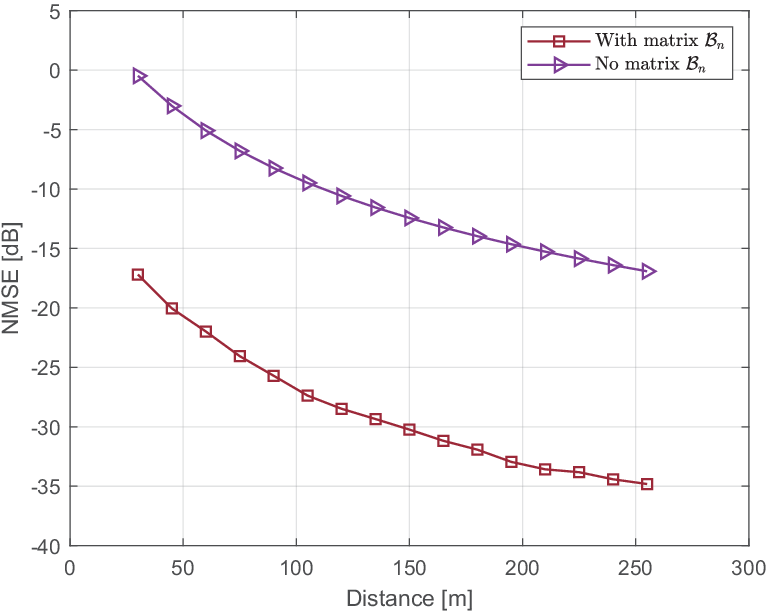}
\caption{The NMSE versus the distances, the BS has 256 antennas, and the UEs move at 120 km/h.}
\label{fig:6}
\end{figure}
In Fig. \ref{fig:6}, we compare the NMSE against the distance to show the advantage of the transform matrix ${{\cal {\bm B}}_n}$, where the distances between the BS and the scatterers increase from 30 m to 255 m.
The curve labelled by ``With matrix ${{\cal {\bm B}}_n}$" is the NMSE when ${{\cal {\bm B}}_n}$ is introduced to transform the spherical wavefront.
We calculate the NMSE by $E\left\{\frac{{\left\| {{{\cal {\bm B}}_n}{{\bf{H}}_{\nu}} - {{\bf{H}}_{{\rm{plane}}}}} \right\|_F^2}}{\left\|{\bf{H}}_{{\rm{plane}}}\right\|_F^2} \right\}$ averaged over UEs.
The other curve is named ``No matrix ${{\cal {\bm B}}_n}$" to show the NMSE between ${{\bf{H}}_{\nu}}$ and ${{\bf{H}}_{\text{plane}}}$, which is calculated by $E\left\{\frac{{\left\| {{{\bf{H}}_{\nu}} - {{\bf{H}}_{{\rm{plane}}}}} \right\|_F^2}}{\left\|{\bf{H}}_{{\rm{plane}}}\right\|_F^2} \right\}$.
The BS antenna configuration is $(1,256,1)$.
We may notice that after introducing ${{\cal {\bm B}}_n}$, the NMSE decreases obviously, and the near-field channel is nearly transformed to a far-field channel.
Therefore, our designed matrix ${{\cal {\bm B}}_n}$ can effectively mitigate the near-field effects.

\begin{figure}[!htb]
\centering
\includegraphics[width=3.3in]{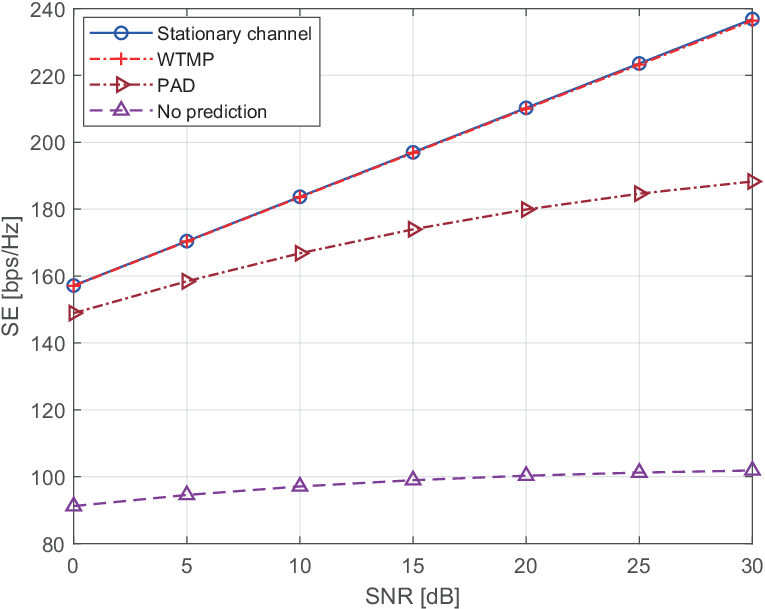}
\caption{The SNR versus SE, the BS has 512 antennas, and the UEs move at 120 km/h.}
\label{fig:7}
\end{figure}
Finally, we adopt a new simulation model consisting of a line-of-sight (LoS) path and 7 clusters. 
Each cluster contains 20 rays, and the total number of propagation paths is 141.
The RMS angular spreads of AOD, EOD, AOA and EOA are updated as $33.2^\circ $, $39.8^\circ $, $140.4^\circ $ and $13.7^\circ $.
Fig. \ref{fig:7} shows the SEs of different prediction methods under this model.
It is clear that our proposed method addresses the near-field effects and time-varying path delay, as the SE of our proposed method is close to the upper bound.

\section{Conclusion}\label{sec:conclusion}
In this paper, we address the mobility problem in ELAA communication systems. 
We propose a wavefront transformation-based near-field channel prediction method by transforming the spherical wavefront. 
We also design a time-frequency-domain projection matrix to capture the time-varying path delay in the mobility scenario, which projects the channel onto the time-frequency domain. 
In the theoretical analysis, we prove that our proposed WTMP method asymptotically converges to be error-free as the number of BS antennas is large enough, given a finite number of samples. 
We also prove that the angles and distance parameters asymptotically determine the designed wavefront-transformation matrix with the increasing number of BS antennas.
Simulation results show that in the high-mobility scenario with large CSI delay, our designed wavefront-transformation matrix provides significant gain, and the performance of our proposed WTMP method is close to the ideal stationary setting.

\appendix

\subsection{Proof of Theorem 1}\label{appendix:Theorem 1}

Providing that observation samples are accurate enough, the projection of the vectorized observation channel ${{{{\bf{\tilde h}}}_{\nu}}(t + {t_\tau })}$ onto the angular-time-frequency basis ${\bf{S}}$ is asymptotically expressed as
\begin{equation}
\begin{array}{l}
\!\!\!\!\!\mathop {\lim }\limits_{{N_t} \to \infty } \frac{{{{\bf{{{\tilde \Omega }}}}_{\nu}}(t+ {t_\tau })}}{{\sqrt {{N_t}} }} = \mathop {\lim }\limits_{{N_t} \to \infty } \frac{{{{\bf{S}}^\dag }{\bf{Q}}{{{\bf{\tilde h}}}_{\nu}}(t+ {t_\tau })}}{{\sqrt {{N_t}} }}\\
= \mathop {\lim }\limits_{{N_t} \to \infty } \frac{{{{\bf{S}}^\dag }{\bf{Q}}{{{\bf{\bar h}}}_{\nu}}(t+ {t_\tau })}}{{\sqrt {{N_t}} }} = \mathop {\lim }\limits_{{N_t} \to \infty } \frac{{{{\bf{S}}^\dag }{{{\bf{\bar h}}}_{{\rm{plane}}}}(t + {t_\tau })}}{{\sqrt {{N_t}} }} + {{\bf{S}}^\dag }{\bf{x}}.
\end{array}
\!\label{projection observation channel}
\end{equation}
Since ${{\bf{s}}_n}$ is the $n$-th column vector of ${\bf{S}}$, the $n$-th row vector of ${\bf{S}}^{\dag}$ is calculated by $\frac{{{{\bf{s}}_n^H}}}{{{N_s^2}}}$. 
Therefore, the $n$-th row element of ${{{\bf{{{\tilde \Omega }}}}_{\nu}}(t+ {t_\tau })}$ is computed by
\begin{equation}
\begin{array}{l}
\mathop {\lim }\limits_{{N_t} \to \infty } \frac{{{\tilde \chi _n}(t+ {t_\tau })}}{{\sqrt {{N_t}} }}
= \mathop {\lim }\limits_{{N_t} \to \infty } \frac{{\sum\limits_{m = 1}^{{N_t}} {} {g_{n,m}}{\tilde \chi _{n,m}}(t+ {t_\tau })}}{{{N_s^2}\sqrt {{N_t}} }}\\
\ \ \ \ \ \ \ \ \ \ \ \ \ \ \ \ \ \ \ = \mathop {\lim }\limits_{{N_t} \to \infty } \frac{{{{\bf{s}}_n^H}{{{\bf{\bar h}}}_{{\rm{plane}}}}(t + {t_\tau })}}{{{N_s^2}\sqrt {{N_t}} }} + \frac{{{{\bf{s}}_n^H}{\bf{x}}}}{{{N_s^2}}}\\
\ \ \ \ \ \ \ \ \ \ \ \ \ \ \ \ \ \ \ = \mathop {\lim }\limits_{{N_t} \to \infty } \frac{{\sum\limits_{m = 1}^{{N_t}} {} {g_{n,m}}{\chi _{n,m}}(t+ {t_\tau })}}{{{N_s^2}\sqrt {{N_t}} }} + \frac{{{{\bf{s}}_n^H}{\bf{x}}}}{{{N_s^2}}},
\end{array}
\!\label{projection observation channel}
\end{equation}
where
\begin{equation}
\begin{array}{l}
\!\!\!\!\!\!{{\tilde \chi}_{n,m}}(t+ {t_\tau }) = \sum\limits_{p = 1}^P {} {\beta _p}{\alpha _{m,p}}{e^{j4\pi {{\tilde \omega }_p}(t+ {t_\tau })}}\\
\!\!\!\!\buildrel \Delta \over = \!\sum\limits_{p = 1}^P {} {\beta _p}{\alpha _{m,p}}{e^{j4\pi{{\omega }_p}(t + {t_\tau })}} \!+\! \Delta x \!=\! {\chi _{n,m}}(t+ {t_\tau }) \!+\! \Delta x,
\end{array}
\!\!\!\label{projection observation channel}
\end{equation}
and $\Delta x$ denotes the error.
Next, we estimate the Doppler by an improved MP method:
Select the first $(N_s-1)$ and the last $(N_s-1)$ samples. Two MP matrices ${{{\bf{\tilde D}}}_1} \in {{\mathbb{C}}^{Q \times (N_s - Q)}}$ and ${{{\bf{\tilde D}}}_2} \in {{\mathbb{C}}^{Q \times (N_s - Q)}}$, may be generated by
\begin{equation}
{{{\bf{\tilde D}}}_1} = {{\bf{E}}_1}{\bf{Z}}{{\bf{R}}_{{t_\tau }}}{{\bf{E}}_2} + \Delta x{{\bf{1}}_{Q \times ({N_s} - Q)}},
\!\label{MP matrix first}
\end{equation}
and
\begin{equation}
{{{\bf{\tilde D}}}_2} = {{\bf{E}}_1}{{\bf{Z}}^2}{{\bf{R}}_{{t_\tau }}}{{\bf{E}}_2} + \Delta x{{\bf{1}}_{Q \times ({N_s} - Q)}},
\!\label{MP matrix last}
\end{equation}
where 
\begin{equation}
\begin{array}{l}
\!\!{{\bf{R}}_{{t_\tau }}} = \\
{\rm{diag}}\{ {\beta _1}{\alpha _{m,1}}{g_{n,m}}{e^{j4\pi{\omega _1}{t_\tau }}}, \cdots , {\beta _P}{\alpha _{m,P}}{g_{n,m}}{e^{j4\pi{\omega _P}{t_\tau }}}\} .
\end{array}
\!\label{MP matrix R}
\end{equation}
Then, define an error matrix $\Delta {\bf{\tilde D}} \in {{\mathbb{C}}^{Q \times (N_s - Q)}}$:
\begin{equation}
\Delta {\bf{\tilde D}} = {{\bf{\tilde D}}_2} - {{\bf{\tilde D}}_1} = {{\bf{E}}_1}{\bf{Z}}\left( {{\bf{Z}} - {\bf{I}}} \right){{\bf{R}}_{{t_\tau }}}{{\bf{E}}_2}.
\!\label{MP matrix error}
\end{equation}
The matrix ${\bf{Z}}$ is estimated by 
\begin{equation}
{\bf{\hat Z}} = eig(\Delta {{{\bf{\tilde D}}}_{2:({N_s} - Q)}}\Delta {\bf{\tilde D}}_{1:({N_s} - Q - 1)}^\dag ),
\!\label{MP matrix Zestimation}
\end{equation}
where $\Delta {{{\bf{\tilde D}}}_{2:({N_s} - Q)}}$ and $\Delta {{{\bf{\tilde D}}}_{1:({N_s} - Q-1)}}$ consist of the first $({N_s} - Q-1)$ and the last $({N_s} - Q-1)$ columns from $\Delta {\bf{\tilde D}}$.
The error matrix $\Delta x{{\bf{1}}_{Q \times ({N_s} - Q)}}$ will not affect the estimation accuracy of ${\bf{\hat Z}}$.
In other words,
\begin{equation}
\mathop {\lim }\limits_{{N_t} \to \infty } \frac{{{{\hat \chi }_n}(t+ {t_\tau }) - {\chi _n}(t+ {t_\tau })}}{{\sqrt {{N_t}} }} = 0.
\!\label{w observation channel}
\end{equation}
We also may rewrite the asymptotic result as
\begin{equation}
\mathop {\lim }\limits_{{N_t} \to \infty } \frac{{\left\| {{{{{\bf{\hat h}}}_{\nu}}(t + {t_\tau })} \!-\! {{{\bf{\bar h}}_{\nu}}(t + {t_\tau })}} \right\|_2^2}}{{{N_t}}} = 0.
\!\label{estimation observation channel}
\end{equation}

Notice that $\mathop {\lim }\limits_{{N_t} \to \infty } \frac{{\left\| {{{\bf{{\bar h}}}_{\nu}}(t + {t_\tau })} \right\|_2^2}}{{{N_t}}} = {\varepsilon _1}$, where ${\varepsilon _1}$ is a positive value.
Then, basing on Eq. (\ref{estimation observation channel}), it might be readily obtained
\begin{equation}
\mathop {\lim }\limits_{{N_t} \to \infty }\!\!\!\!\!\! \frac{{\left\| {{{{{\bf{\hat h}}}_{\nu}}(t + {t_\tau })} \!-\! {{{\bf{\bar h}}_{\nu}}(t + {t_\tau })}} \right\|_2^2}}{{\left\| {{{{\bf{\bar h}}_{\nu}}(t + {t_\tau })}} \right\|_2^2}} = 0.
\!\label{theorem1 proof}
\end{equation}
Thus, Theorem \ref{theorem} is proved.

\ifCLASSOPTIONcaptionsoff
  \newpage
\fi

\bibliographystyle{IEEEtran}
\bibliography{references.bbl}

\begin{thebibliography}{10}
\providecommand{\url}[1]{#1}
\csname url@samestyle\endcsname
\providecommand{\newblock}{\relax}
\providecommand{\bibinfo}[2]{#2}
\providecommand{\BIBentrySTDinterwordspacing}{\spaceskip=0pt\relax}
\providecommand{\BIBentryALTinterwordstretchfactor}{4}
\providecommand{\BIBentryALTinterwordspacing}{\spaceskip=\fontdimen2\font plus
\BIBentryALTinterwordstretchfactor\fontdimen3\font minus \fontdimen4\font\relax}
\providecommand{\BIBforeignlanguage}[2]{{%
\expandafter\ifx\csname l@#1\endcsname\relax
\typeout{** WARNING: IEEEtran.bst: No hyphenation pattern has been}%
\typeout{** loaded for the language `#1'. Using the pattern for}%
\typeout{** the default language instead.}%
\else
\language=\csname l@#1\endcsname
\fi
#2}}
\providecommand{\BIBdecl}{\relax}
\BIBdecl

\bibitem{Marzetta10TCom}
T.~L. {Marzetta}, ``Noncooperative cellular wireless with unlimited numbers of base station antennas,'' \emph{IEEE Trans. Wireless Commun.}, vol.~9, no.~11, pp. 3590--3600, Nov. 2010.

\bibitem{Cheng19TWC}
X.~{Cheng}, K.~{Xu}, J.~{Sun}, and S.~{Li}, ``Adaptive grouping sparse {Bayesian} learning for channel estimation in non-stationary uplink massive {MIMO} systems,'' \emph{IEEE Trans. Wireless Commun.}, vol.~18, no.~8, pp. 4184--4198, Aug. 2019.

\bibitem{Wang23TVT}
L.~{Wang}, B.~{Ai}, Y.~{Niu}, H.~{Jiang}, S.~{Mao}, Z.~{Zhong}, and N.~{Wang}, ``Joint user association and transmission scheduling in integrated mmwave access and terahertz backhaul networks,'' \emph{IEEE Trans. Veh. Technol.}, pp. 1--11, 2023.

\bibitem{Huang19TWC}
C.~{Huang}, A.~{Zappone}, G.~C. {Alexandropoulos}, M.~{Debbah}, and C.~{Yuen}, ``Reconfigurable intelligent surfaces for energy efficiency in wireless communication,'' \emph{IEEE Trans. Wireless Commun.}, vol.~18, no.~8, pp. 4157--4170, Aug. 2019.

\bibitem{Podkurkov21TSP}
I.~Podkurkov, G.~Seidl, L.~Khamidullina, A.~Nadeev, and M.~Haardt, ``Tensor-based near-field localization using massive antenna arrays,'' \emph{IEEE Trans. Signal Process.}, vol.~69, pp. 5830--5845, Aug. 2021.

\bibitem{Lu23TCom}
Y.~{Lu} and L.~{Dai}, ``Near-field channel estimation in mixed {LoS/NLoS} environments for extremely large-scale {MIMO} systems,'' \emph{IEEE Trans. Commun.}, vol.~71, no.~6, pp. 3694--3707, June 2023.

\bibitem{Molisch07}
A.~F. {Molisch}, \emph{Wireless communications}.\hskip 1em plus 0.5em minus 0.4em\relax John Wiley \& Sons, 2007.

\bibitem{Lu22TWC}
H.~{Lu} and Y.~{Zeng}, ``Communicating with extremely large-scale array/surface: {Unified} modeling and performance analysis,'' \emph{IEEE Trans. Wireless Commun.}, vol.~21, no.~6, pp. 4039--4053, June 2022.

\bibitem{Yang21TWC}
J.~{Yang}, Y.~{Zeng}, S.~{Jin}, and C.~K. {Wen}, ``Communication and localization with extremely large lens antenna array,'' \emph{IEEE Trans. Wireless Commun.}, vol.~20, no.~5, pp. 3031--3048, May 2021.

\bibitem{21TSPBayesian}
Y.~{Zhu}, H.~{Guo}, and V.~K.~N. {Lau}, ``Bayesian channel estimation in multi-user massive {MIMO} with extremely large antenna array,'' \emph{IEEE Trans. Signal Process.}, vol.~69, pp. 5463--5478, Sept. 2021.

\bibitem{20WCLChannel}
Y.~{Han}, S.~{Jin}, C.~K. {Wen}, and X.~{Ma}, ``Channel estimation for extremely large-scale massive {MIMO} systems,'' \emph{IEEE Wireless Commun. Lett.}, vol.~9, no.~5, pp. 633--637, May 2020.

\bibitem{Magoarou19}
L.~L. {Magoarou}, A.~L. {Calvez}, and S.~{Paquelet}, ``Massive {MIMO} channel estimation taking into account spherical waves,'' in \emph{Proc. IEEE Int. Workshop Signal Process. Adv. Wireless Commun. (SPAWC)}, Cannes, France, July 2019, pp. 1--5.

\bibitem{96Validity}
G.~W. {Forbes}, ``Validity of the fresnel approximation in the diffraction of collimated beams,'' \emph{J. Opt. Soc. Am. A}, vol.~13, no.~9, pp. 1816--1826, Sept. 1996.

\bibitem{19TAP}
Z.~{Zheng}, M.~{Fu}, W.~Q. {Wang}, S.~{Zhang}, and Y.~{Liao}, ``Localization of mixed near-field and far-field sources using symmetric double-nested arrays,'' \emph{IEEE Trans. Antennas Propag.}, vol.~67, no.~11, pp. 7059--7070, Nov. 2019.

\bibitem{22RIS}
Y.~{Pan}, C.~{Pan}, S.~{Jin}, and J.~{Wang}, ``{RIS}-aided near-field localization and channel estimation for the terahertz system,'' \emph{IEEE J. Sel. Topics Signal Process.}, pp. 1--14, June 2023.

\bibitem{20TVT}
Y.~{Cao}, T.~{Lv}, Z.~{Lin}, P.~{Huang}, and F.~{Lin}, ``Complex resnet aided {DoA} estimation for near-field {MIMO} systems,'' \emph{IEEE Trans. Veh. Technol.}, vol.~69, no.~10, pp. 11\,139--11\,151, July 2020.

\bibitem{23WCL}
J.~{Xiao}, J.~{Wang}, Z.~{Chen}, and G.~{Huang}, ``{U-MLP}-based hybrid-field channel estimation for {XL-RIS} assisted millimeter-wave {MIMO} systems,'' \emph{IEEE Wireless Commun. Lett.}, vol.~12, no.~6, pp. 1042--1046, Mar. 2023.

\bibitem{22TComDai}
M.~{Cui} and L.~{Dai}, ``Channel estimation for extremely large-scale {MIMO}: Far-field or near-field?'' \emph{IEEE Trans. Commun.}, vol.~70, no.~4, pp. 2663--2677, Jan. 2022.

\bibitem{Yin20JSAC}
H.~{Yin}, H.~{Wang}, Y.~{Liu}, and D.~{Gesbert}, ``Addressing the curse of mobility in massive {MIMO} with prony-based angular-delay domain channel predictions,'' \emph{IEEE J. Sel. Areas Commun.}, vol.~38, no.~12, pp. 2903--2917, Dec. 2020.

\bibitem{21GaoJSAC}
H.~{Jiang}, M.~{Cui}, D.~W.~K. {Ng}, and L.~{Dai}, ``Accurate channel prediction based on transformer: {Making} mobility negligible,'' \emph{IEEE J. Sel. Areas Commun.}, vol.~40, no.~9, pp. 2717--2732, Sept. 2022.

\bibitem{12TomasTVT}
T.~{Zemen}, L.~{Bernado}, N.~{Czink}, and A.~F. {Molisch}, ``Iterative time-variant channel estimation for 802.11p using generalized discrete prolate spheroidal sequences,'' \emph{IEEE Trans. Veh. Technol.}, vol.~61, no.~3, pp. 1222--1233, Mar. 2012.

\bibitem{3GPPR16}
3GPP, \emph{Study on channel model for frequencies from 0.5 to 100 {GHz} ({Release} 16)}.\hskip 1em plus 0.5em minus 0.4em\relax Technical Report TR 38.901, available: http://www.3gpp.org.

\bibitem{18Conference}
D.~{Dardari} and F.~{Guidi}, ``Direct position estimation from wavefront curvature with single antenna array,'' in \emph{Proc. 8th Int. Conf. on Localization and GNSS (ICL-GNSS)}, Guimaraes, Portugal, Nov. 2018, pp. 1--5.

\bibitem{MDMP}
W.~{Li}, H.~{Yin}, Z.~{Qin}, Y.~{Cao}, and M.~{Debbah}, ``A multi-dimensional matrix pencil-based channel prediction method for massive {MIMO} with mobility,'' \emph{IEEE Trans. Wireless Commun.}, vol.~22, no.~4, pp. 2215--2230, Apr. 2023.

\bibitem{Yin16TSP}
H.~{Yin}, L.~{Cottatellucci}, D.~{Gesbert}, R.~R. {Müller}, and G.~{He}, ``Robust pilot decontamination based on joint angle and power domain discrimination,'' \emph{IEEE Trans. Signal Process.}, vol.~64, no.~11, pp. 2990--3003, Feb. 2016.

\bibitem{EZF}
L.~{Sun} and M.~R. {McKay}, ``Eigen-based transceivers for the {MIMO} broadcast channel with semi-orthogonal user selection,'' \emph{IEEE Trans. Signal Process.}, vol.~58, no.~10, pp. 5246--5261, Oct. 2010.

\end{thebibliography}

\end{document}